\documentclass[sigconf,screen]{acmart}

\usepackage{fontawesome5}

\usepackage{amsmath,amssymb,amsfonts}
\usepackage{subcaption}
\usepackage{bbm,epsfig,bm}
\usepackage{algorithmic}
\usepackage{graphicx}
\usepackage{textcomp}
\usepackage{dcolumn}
\usepackage{xspace}
\usepackage[normalem]{ulem}
\usepackage[binary-units=true]{siunitx}
\usepackage{booktabs}
\usepackage{arydshln} 
\usepackage[capitalise]{cleveref}
\usepackage{tcolorbox}
\usepackage{multicol}
\usepackage{wrapfig}
\usepackage{framed}
\usepackage{caption}
\usepackage[T1]{fontenc}

\usepackage{threeparttable}
\usepackage{transparent}

\newcommand{\eg}{e.g.,\xspace}
\newcommand{\ie}{i.e.,\xspace}

\newcounter{HCounter}
\newcommand{\Hyp}[2]{%
\refstepcounter{HCounter} \label{#1}
    \smallskip
	\noindent 
	\textbf{H}$_{\arabic{HCounter}}$.~\emph{#2}
    \smallskip
}
\newcommand{\hr}[1]{\textbf{H}$_{\ref{#1}}$}

\newcommand{\mysec}[1]{\vspace{0.1cm} \noindent \textbf{#1.}}
\newcommand{\mysubsec}[1]{\vspace{0.2cm} \noindent \textit{#1}}

\usepackage{array}
\newcommand{\PreserveBackslash}[1]{\let\temp=\\#1\let\\=\temp}
\newcolumntype{P}[1]{>{\raggedright\arraybackslash}p{#1}}
\newcolumntype{C}[1]{>{\PreserveBackslash\centering}p{#1}}
\newcolumntype{R}[1]{>{\PreserveBackslash\raggedleft}p{#1}}
\newcolumntype{L}[1]{>{\PreserveBackslash\raggedright}p{#1}}

\setcopyright{none} 
\acmConference[Draft]{}{}{}
\renewcommand\footnotetextcopyrightpermission[1]{}

\begin{document}

\title[The Strength of Weak Ties Between Open-Source Developers]{The Strength of Weak Ties Between Open-Source Developers}
\subtitle{In the Python Ecosystem GitHub Stars Predict Innovation Better Than Commits}

\DeclareRobustCommand*{\AuthorRefMark}[1]{\raisebox{0pt}[0pt][0pt]{\textsuperscript{\footnotesize%
    \ifcase#1\or \faStar[solid]\or \faStar[regular] \else\textsuperscript{\expandafter\romannumeral#1}\fi}}}

\author{Hongbo Fang,\AuthorRefMark{1}\AuthorRefMark{2} Patrick Park,\AuthorRefMark{1} James Evans,\AuthorRefMark{2} James Herbsleb,\AuthorRefMark{1} and Bogdan Vasilescu\AuthorRefMark{1}}
\affiliation{
  \vspace{2pt}
  \institution{\normalsize \AuthorRefMark{1}Carnegie Mellon University \hspace{0.3em} \AuthorRefMark{2}University of Chicago}
  \country{}
}
\email{{hongbofang, jevans}@uchicago.edu, {patpark, jim.herbsleb, vasilescu}@cmu.edu}

\renewcommand\shortauthors{Hongbo Fang, Patrick Park, James Evans, James Herbsleb, Bogdan Vasilescu}

\begin{abstract}
In a real-world social network, weak ties (reflecting low-intensity, infrequent interactions) act as bridges and connect people to different social circles, giving them access to diverse information and opportunities that are not available within one’s immediate, close-knit vicinity. Weak ties can be crucial for creativity and innovation, as they introduce ideas and approaches that people can then combine in novel ways, leading to innovative solutions. Do weak ties facilitate creativity in software in similar ways?
This paper suggests that the answer is ``yes.'' Concretely, we study the correlation between developers' knowledge acquisition through three distinct interaction networks on GitHub and the innovativeness of the projects they develop, across over 37,000 Python projects hosted on GitHub. 
Our findings suggest that the topical diversity of projects in which developers engage, rather than the volume, correlates positively with the innovativeness of their future code.
Notably, exposure through weak interactions (e.g., starring) emerges as a stronger predictor of future novelty than via strong ones (e.g., committing).

\end{abstract}

\maketitle

\section{Introduction}
\label{sec:intro}


Think of examples of big software innovations. You might name the Netscape browser (created the visual web and fueled mass internet adoption), the Git version control system (revolutionized collaborative software development), Hadoop MapReduce (enabled ``big data'' processing), Photoshop (made digital image manipulation mainstream), Netflix (killed video rental stores), Uber (``revolutionized'' transportation),
or perhaps more than anything else, the World Wide Web (the global information network that connected humanity and fundamentally changed how we access knowledge).

Besides their undeniable impact, what all these innovations have in common is, perhaps surprisingly, that \textit{none was highly novel in a pure sense}.  Instead, they succeeded by recognizing latent potential in existing components that others had overlooked or dismissed as insufficient.
This reveals a fundamental paradox in software innovation: breakthroughs can emerge not only from inventing entirely new concepts, but from developing the insight to see how familiar pieces can fit together in unexpected ways. 

As he remarkably candidly notes in his book ``Weaving the Web''~\citep{berners1999weaving}, Berners-Lee didn't invent hypertext, markup languages, or networking protocols, but he saw how HTTP, HTML, and URLs could create a self-reinforcing web that previous hypertext systems had failed to achieve. 
The web browsers didn't invent hypertext or networking either, but they made the connection between them obvious in retrospect. 
Git didn't create new concepts around file differencing, cryptographic hashing, or distributed systems, but made distributed version control finally practical. 
Photoshop didn't invent image processing algorithms or digital manipulation, but unified traditional darkroom techniques with mathematical transforms in an intuitive interface. 
Uber didn't invent transportation technology, but cleverly orchestrated GPS, mobile payments, and dispatch algorithms that had existed separately for years. 
Finally, Netflix didn't invent video streaming or recommendation systems -- they combined content delivery networks with collaborative filtering at a moment when bandwidth made it viable.

\begin{figure}[t]
\centering
\includegraphics[width=\columnwidth, clip=true, trim=0 0 0 20]{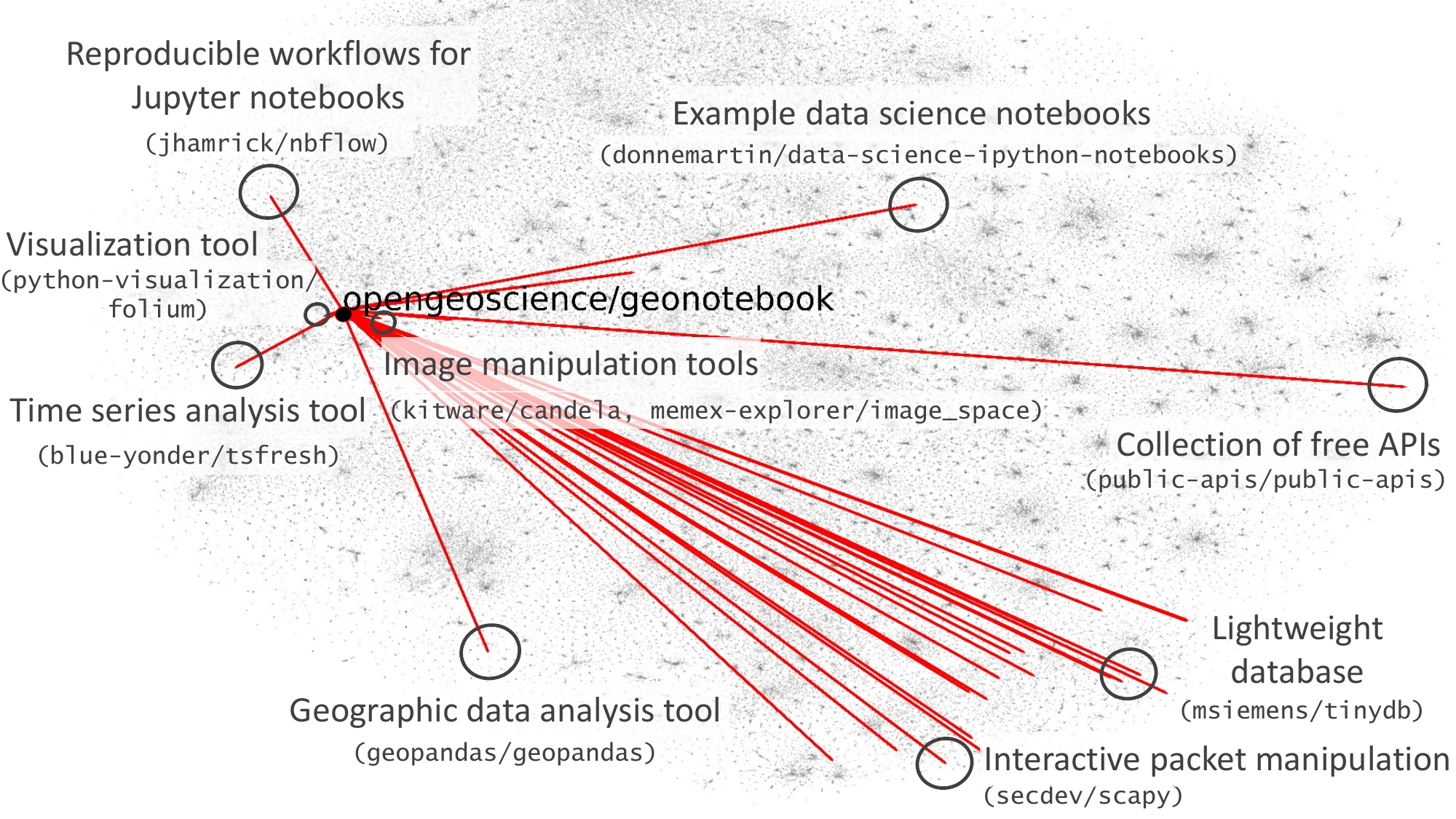}\vspace{-0.2cm}
\caption{t-SNE visualization of the embedding space for weak ties (details in Section~\ref{sec:network-variables}), depicting all the weak ties of the \href{https://github.com/opengeoscience/geonotebook}{GeoNotebook} Python project in our sample. We highlight some that seem influential for the design of the focal project.}
\label{fig:geonotebook}
\end{figure}

What distinguishes many impactful software innovations, then, isn't pure novelty but rather a form of \textit{combinational creativity}~\cite{boden2004creative} -- the ability to perceive new relationships between existing elements and to pursue combinations that may seem counterintuitive. Many innovators aren't necessarily the first to create the individual components, but they are the first to see the gestalt that emerges from their combination.
This suggests that thinking of breakthrough innovations as pure inventions may actually mislead us about where a lot of transformational software comes from. 
Perhaps a highly valuable skill isn't creating something from nothing, but developing the pattern recognition to spot which existing pieces, when properly orchestrated, can produce emergent behaviors that transcend their individual capabilities.
Much real innovation, thus, lies not in the novelty of the parts, but in the non-obvious wisdom of their assembly.
In fact, this process of innovation through novel recombination is not specific to software --
it appears to be a fundamental pattern in virtually all domains of human innovation, and it has been theorized and studied empirically in many contexts, including business, medicine, science, and technology~\cite{schumpeter1939business, uzzi2013atypical, hofstra2020diversity, shi2023surprising}.

But while it's clear that innovation drives the software industry~\cite{edison2013towards}, and that great software engineers (among many other attributes) should be ``\textit{[able to] generate novel and innovative solutions based on the context and its limitations}''~\cite{li2020distinguishes}, it's still unclear how to develop one's ability to combine existing things in clever ways.
Such combinational creativity ``\textit{typically requires a very rich store of knowledge, of many different kinds, and the ability to form links of many different types}''~\cite{boden2004creative}.
How to develop these?
Although empirical evidence is scarce~\cite{mohanani2017perceptions, groeneveld2021exploring}, one might theorize that mixing knowledge depth with knowledge breadth is the key~\cite{da2015niche, turner2002exploring}, since these characteristics make software professionals more successful in general~\cite{sonnentag1995excellent, dieste2017empirical}.
According to this thinking, deep engagement with the software development process, ideally through ``deliberate practice''~\cite{baltes2018towards}, should help build expertise, while broad exploration in many different contexts should help build vocabulary.

Our paper goes one step further in this line of thinking, revealing a stronger and more nuanced effect of breadth than one might expect. 
Across over 37,000 Python projects on GitHub, while we confirm an association between knowledge breadth and software innovation, we find that \textbf{exposure through minimal-effort interactions like starring repositories emerges as a stronger predictor of future innovation than high-effort engagements like writing code}. 
This counterintuitive result suggests that casual, passive observation of diverse projects -- essentially \textit{digital lurking} -- may be more valuable for sparking creative recombination than direct, intensive engagement and active learning.

\looseness=-1
Much like the unexpected advantage of weak ties in job searches, which challenged conventional wisdom of social networks at the time~\cite{granovetter1973strength}, this ``strength of weak ties on GitHub'' represents a paradox worth exploring: 
Sometimes, developers’ least-effort interactions with projects may contribute the most to their innovation capacity. 
Our research investigates this surprising mechanism, examining how the structural characteristics of developers' interaction networks correlate with the innovativeness of the software artifacts they subsequently create.

Below, we draw from established theory 
to formulate our hypotheses (\S\ref{sec:theory}); propose a novel software innovativeness measure, reconstruct strong- and weak-tie interaction networks (via commits, issues, and starring) for a project's core team, and estimate the amount and diversity of knowledge accessible to the core developers via these networks 
(\S\ref{sec:methods}); build regression models to test the association between network structural properties and software innovativeness (\S\ref{sec:results}); and discuss the implications of our results (\S\ref{sec: discussion}).

\section{Theory and Hypotheses}
\label{sec:theory}


In a nutshell, developers on the GitHub platform engage in a wide range of interactions, corresponding to ties of varying strength in a network sense, that create opportunities for knowledge acquisition via social learning. 
This knowledge can influence the innovativeness of the solutions they create.
Before diving into the details of the mechanism, let us clarify some key constructs.

\mysec{Innovation as Novel Recombination}
As discussed above, we view software innovation as often emerging from \textit{novel recombinations} of existing components, libraries, and patterns, 
following the long-standing view that ``[business] innovation combines factors in a new way, or that it consists in carrying out new combinations''~\cite{schumpeter1939business}, which has fueled much research in the social sciences (e.g.,~\cite{uzzi2013atypical, shi2023surprising, hofstra2020diversity}). 
%
As a result, the line between \textit{novelty} and \textit{innovation} becomes fuzzy, thus we will use the two terms interchangeably, in contrast to some prior work that distinguishes them~\cite{couger1992measurement, beghelli2020novelty}.

\mysec{Creativity as Antecedent of Innovation}
How to facilitate the emergence of innovation is unclear, although a common argument centers on enabling creativity as a precursor. 
\textit{Creativity} and \textit{innovation} are closely related but distinct constructs: creativity is the ability to generate novel ideas, while innovation transforms novel concepts into tangible outcomes.
It is expected that a software team that is more creative has access to richer knowledge (\eg knowledge about the problem and technical approaches to solving it), and is better able to 
combine pieces of that knowledge to create an innovative solution~\cite{du2007role}. 
Of course, there are many facets of creativity, including as an individual personality trait~\cite{kandler2016nature}.
In this paper we focus only on one. Specifically, we argue that creativity at the team level is a function of the knowledge networks of the team members and, while we cannot measure creativity directly, we can expect to see differences in the teams' creative outputs (\ie more novel software), that are associated with differences in the structure of those knowledge networks.

\mysec{Knowledge Acquisition via Social Learning}
%
Historically, and especially with the advent of ``social coding'' platforms like GitHub, participation in open source has been rife with opportunities for social learning~\cite{crowston2005social}, \ie the process by which individuals acquire new knowledge, behaviors, skills, or attitudes through observation, imitation, and interaction with others.
Learning, including by observing what others are doing~\cite{yang2021differential}, has been~\cite{lakhani2005hackers, krishnamurthy2006intrinsic} and remains~\cite{gerosa2021shifting} among the most important motivations for people to contribute to open source. 
In addition, modern code hosting platforms offer many opportunities for users to interact (\eg collaboration on a shared codebase, issue discussions, code review).
Through social media-like functionality (\eg ``following''~\cite{blincoe2016understanding}) and many available signals that offer a high degree of transparency~\cite{dabbish2012social} (\eg repository badges~\cite{trockman2018adding}), the platforms also facilitate users quietly observing and being influenced by others' behaviors.

These interactions with other individuals~\cite{von2003community} and with the artifacts they create~\cite{cervetti2009increasing} serve as channels for information dissemination and knowledge exchange. 
Indeed, prior studies have documented many such social learning effects, including choosing an open-source license~\cite{peng2013learning, singh2013networks}, discovering and adopting emerging tools~\cite{singer2014software, lamba2020heard}, and learning new software design principles and programming skills by reviewing code authored by others~\cite{nandigam2008learning, alami2019does}.

Therefore, we expect that an increase in the volume of interactions among developers should correspond to a higher amount of knowledge transfer, thus better preparing developers for future innovations. Thus, we hypothesize that:

\Hyp{hyp1}{The more interactions developers have with other developers and projects, the more innovative their projects are.}

Still, if innovation requires an increased ``vocabulary'' of knowledge bits accessible  for recombination, even high volumes of interaction might not be sufficient if the corresponding information is redundant. 
This can happen, \eg when one 
works primarily within a narrow domain.
Instead, we expect that accessing diverse knowledge acts as a catalyst for innovation, enabling individuals to integrate disparate concepts and to develop unconventional and novel products~\cite{schumpeter2021theory}.
Outside of software engineering, \citet{tortoriello2015being} observed a positive association between employees' access to diverse knowledge within the research departments of high-tech companies and their innovation levels. 
Similarly, \citet{abdul2019diversity} found that the integration of knowledge from both internal research teams and customers in the public sector enhances innovation.
Thus, we hypothesize that:

\Hyp{hyp2}{The greater the informational diversity of developers' past interactions,
the more innovative their projects are.}

\mysec{The Strength of Weak Ties}
In a social network, ties can have varying strength, reflective of real-world factors like the duration of shared interactions, emotional depth, level of intimacy, and amount of reciprocal exchanges~\cite{granovetter1973strength}.
In that sense, ``weak'' ties involve infrequent and less intimate interactions, while ``strong'' ties are characterized by higher frequency of interaction and intimacy.

In the late 1960s, sociologist Mark Granovetter uncovered a counterintuitive finding that would revolutionize our understanding of social networks: When searching for jobs, people relied more on casual acquaintances (weak ties) than close friends or family (strong ties). 
His groundbreaking theory on the ``strength of weak ties''~\cite{granovetter1973strength} revealed that these seemingly superficial connections often serve as critical bridges between different social circles, providing access to novel information unavailable within one’s immediate network.
This pattern arises because individuals within the same network neighborhood have numerous opportunities for interaction and often share many mutual ties. This shared social context leads to increased similarity in behaviors, interests, and other characteristics.
Consequently, the information disseminated via strong ties risks becoming less novel and less valuable (the discussion of echo chambers on social media~\cite{kitchens2020understanding} is a prime example of this degradation of information quality).
In contrast, information acquired via weak ties, or cross-group connections, tends to originate more from individuals with diverse backgrounds and knowledge bases, making it more likely to be novel and of greater value~\cite{granovetter1973strength}.
This mechanism has been extensively validated in social contexts, including generation of innovative ideas~\cite{burt2004structural}, the diffusion of news feed content~\cite{bakshy2012role}, and job seeking in the digital age~\cite{rajkumar2022causal}.

Similarly, we expect that the software-related knowledge transferred through weak ties is more valuable for the creation of innovative software projects. 
Thus, we hypothesize that:

\Hyp{hyp3}{
The more the informational diversity of developers' past interactions is due to weak ties, the more innovative their projects are.
}

\section{Methods}
\label{sec:methods}

Next we give a high-level overview of our approach, before diving into operationalization details.

\subsection{Overview / Intuition}

\mysec{Network Construction}
We theorized above that (1)~the interactions developers have with each other and with each other's artifacts, \ie the ties they form, create opportunities for knowledge transfer and facilitate social learning; and (2)~``strong'' and ``weak'' ties may play different roles in this knowledge diffusion process.
To operationalize these concepts, we first construct three interaction networks where every node is a project (\ie GitHub repository), and the links between two nodes encode different interactions the developers of one project had with the other project. 

There are many ways in which two open-source developers can interact, both in and outside the GitHub platform. 
Clearly, it's not possible to capture all interactions at scale, as projects may use, \eg a diversity of communication channels, including private ones.
Instead, we consider three representative examples of interactions that (a)~vary substantially in effort, thus can be expected to reflect varying levels of knowledge flow from the target project to the author of the action (\ie to encode person--to--project ties of varying strength, which we later project to construct our project--to--project networks), (b)~are core features of the GitHub platform, thus are commonly used:
making \textit{commits} to a codebase (direct pushes or merged pull requests), posting or participating in \textit{issue discussions}, and \textit{starring} repositories.

Among the three, commits typically require the most effort and starring the least.
This ranking should correspond to the depth of the knowledge that may transfer as a result of each action.
Intuitively, while commits vary in size and content~\cite{alali2008s}, an ``average'' commit should require a significant understanding of the project's technical details~\cite{von2003community}, indicating a path for considerable knowledge flow between the project and the commit author.
Issue threads primarily discuss feature suggestions, bug reports, and user support~\cite{merten2015requirements}. 
Thus, while engaging in issue discussions can reflect a deep understanding of the project, on average we expect that it requires (and reflects) a less deep understanding of the project compared to making changes to its codebase. 
Finally, starring a project is often done as a sign of appreciation, a bookmarking attempt, or an intent for later use~\cite{borges2018s}. 
Starring a repository indicates at least some awareness of the project, but on average probably much more basic understanding than the other two. 

Note that we consider only the actions of core developers in a focal project, identified heuristically as those contributors who authored at least five
percent of all commits, with a minimum of 10 commits in total. Our operationzalization is validated with alternative threshold of core developer identification in the appendix.
Conceptually, in open source, core developers wield the most influence over a project's technical decisions and development trajectory~\cite{robles2009evolution}. 
Empirically, in our sample core developers also authored the vast majority (close to 90 percent) of commits importing new packages into projects (which we use to compute innovativeness, as described below).
Thus, we expect that the knowledge they had access to prior to working on a focal project (as opposed to peripheral contributors) 
has the most influence on the design and innovativeness of that project.

As possible alternatives~\cite{robillard2024communicating}, we considered reflecting the past interactions of only the founder of a project (which would have missed influential people joining a project later), as well as network centrality-based operationalizations of ``core'' versus ``periphery'' status (which are more precise but computationally more complex and, on average, correlate highly with count-based measures like ours~\cite{joblin2017classifying}). 
Finally, we considered capturing the past interactions of all project contributors. 
However, at the scale of our study (over 37,000 projects) this was infeasible considering that popular projects may have tens of thousands of peripheral contributors.

\mysec{Network Measures}
Using our network data, we compute variables related to the network position of each project to represent the \textit{amount} and \textit{diversity of knowledge} that may be accessible to its core developers through each network. 
For example, consider the commit-based project--to--project network.
A node (project) in this network may have many outgoing connections, indicating that its core developers have collectively committed to many other projects prior to joining the focal project.
That is, they have first-hand experience with the technologies used in those other projects, and they can probably draw from that experience (knowledge) now to influence the development of the focal project.
Similarly, we can reason about the other two networks constructed from issue discussions and starring repositories.

How to measure the diversity of knowledge is less obvious.
Conceptually, the amount and diversity of knowledge are related yet distinct. 
For example, a project whose core developers are connected to a large number of other projects within the network \textit{may} have access to diverse knowledge. 
However, the actual diversity of this knowledge remains contingent on the nature of these connections -- if all linked projects contain similar knowledge, diversity may be limited; e.g., if they're all related to visualization, that doesn't say much about experience with \textsc{tensorflow}. 


To capture this diversity, we adopt an approach analogous to word embeddings, wherein semantic similarity between words is inferred from their vector representations~\cite{mikolov2013distributed}.
The key ideas is that we compute project graph embeddings and use embedding distance to quantify the similarity of gainable knowledge among projects. 
The intuition is that nodes that frequently appear together in the same random walk while learning the embeddings represent projects that are often contributed to or interacted with by the same developers, \ie are related in some informational sense.
This could be because they share similar functionality, are used in similar contexts, or are part of the same development ecosystem.

\mysec{Dimensionality Reduction}
Thus far we have been treating our three networks (commits, issues, and stars) as separate. 
However, one can expect that the variables we compute based on these networks are correlated to some extent. 
Moreover, one can imagine considering other types of interactions (\ie networks) besides our three, which would result in yet more variables.
To prevent issues of multicollinearity in our subsequent regressions, which can both complicate the interpretation of the estimated coefficients and reduce statistical power~\cite{daoud2017multicollinearity}, we use Principal Component Analysis (PCA)~\cite{wold1987principal} to decompose the variables representing the \textit{amount} and \textit{diversity of knowledge} into orthogonal components.
As we show below, this produces a clear decomposition of our original variables into two components capturing strong and weak ties, which we use in our models instead of the original variables.

\mysec{Project Innovativeness Measure}
\citet{fang2024novelty} introduced a measure of a software project's novelty, or innovativeness, as a function of the combinations of packages the project imports.
For example, while certain libraries, such as \textsc{numpy} and \textsc{tensorflow}, are frequently used together, others, like \textsc{numpy} and \textsc{requests}, are less commonly combined. The measure estimates how atypical combining two packages is, and by extension how atypical is the overall set of packages being used as dependencies in a project, by comparison to what could be expected by random chance. 
Projects that import more atypically paired packages are deemed more innovative as they reuse packages in nontraditional ways.

We maintain this framing of \textbf{innovation through novel recombination}, and similarly consider \textbf{packages as the unit of recombination}.
Packages are designed to be reused; they're typically substantial enough to provide real functionality while remaining composable (individual functions are too small -- they lack sufficient context and capability to be meaningfully recombined; entire applications are too large); 
and they're usually semantically coherent, i.e., they tend to be organized around conceptual domains (e.g., \textsc{requests} encapsulates the entire complexity of HTTP communication in a reusable form, while \textsc{pandas} bundles decades of data manipulation knowledge into a coherent interface).
Alternatively, one could consider something like ``concepts'' as the unit of recombination, as in some prior sience-of-science work~\citep{hofstra2020diversity}. 
However, it wasn't clear how to automatically extract concepts from software, so we chose packages as the unit instead.

Our measure of project innovativeness, while similar in spirit to Fang et al's~\citep{fang2024novelty}, is based on learning embeddings of packages and addresses two important limitations of the original measure.
First, the original measure only captures pairwise combinations between packages (and aggregates up to the project level from that), neglecting the possible interactions among related packages. 
However, packages often form ``stacks'' comprised of more than two libraries that complement each other and are often reused together. 
Our embeddings-based approach considers more context and is expected to better capture a package's technical use, function, and ``stack.''
Second, computing the original measure requires random reshuffling of the global dependency network, which is computationally prohibitive at the scale of our study. 
In contrast, we compute the cosine similarity of embeddings, which is faster.

\mysec{Regression Analysis}
Putting everything we discussed so far together, our analysis involves modeling the variation in project innovativeness across our sample as a function of the variables we discussed above, controlling for known covariates.
While not causal,\footnote{Testing our hypotheses experimentally is infeasible, and identification is otherwise unclear from  observational data like ours, since there is no clear ``intervention.''} this analysis will still allow us to test the extent to which the observational data is consistent with the causal paths theorized above (about determinants of combinatorial creativity).

In the remainder of this section we give lower-level operationalization details for the steps above, including validation checks.

\subsection{Sample Selection}
\label{subsec:dataset}

\looseness=-1
We combine data from World of Code (WoC)~\cite{ma2019world} and the GitHub API. 
WoC is arguably the most comprehensive record of open-source projects and their commit history, and it also contains historical import-based dependency information for projects in the most popular programming languages, which we use when computing project innovativeness.
%
To keep our analysis tractable, we focus only on Python projects, \ie the project contains more than 10 Python source code scripts. 
We chose Python due to its immense popularity~\cite{srinath2017python} and its widespread use across various domains by developers from diverse backgrounds~\cite{saabith2019python}. 
While not allowing us to directly generalize our findings, the considerable variability in the Python ecosystem across many characteristics should ensure points of overlap with many other ecosystems.

Because many public repositories are intended as code dumps rather than active development projects, and to account for more of the intricacies of GitHub data~\cite{kalliamvakou2016depth}, we excluded forks and repositories with fewer than ten commits in total. 
In addition, we excluded repositories created before 2008 (pre GitHub) and after 2022 (which allows for at least one year of history in WoC, which we need to make inferences about project innovativeness). 

For this remaining set of projects we queried the GitHub API for historical information on all their commits, issues, and stars (the latter two are not available in WoC); for each such event we recorded its author and timestamp. 
We used this data for our network construction, detailed in Section~\ref{sec:network-construction} below.

As a final preprocessing step, we heuristically identified and filtered out bot accounts~\cite{dey2020detecting, wessel2018power}, which would distort our networks and subsequent measures, as they tend to have a lot of activity. 
To this end, we searched for keywords in user logins, such as \textit{-bot} or \textit{-robot}, and labeled those users as bots after manual review. 
Additionally, we manually inspected the profile descriptions of the top 100 most active user accounts in our sample, ordered by the number of commits, to ensure the removal of highly active bot accounts that may have escaped the naming convention heuristics.

\subsection{Network Construction}
\label{sec:network-construction}

\begin{figure}[t]
\centering
\includegraphics[width=0.95\columnwidth, clip=true, trim=0 340 300 0]{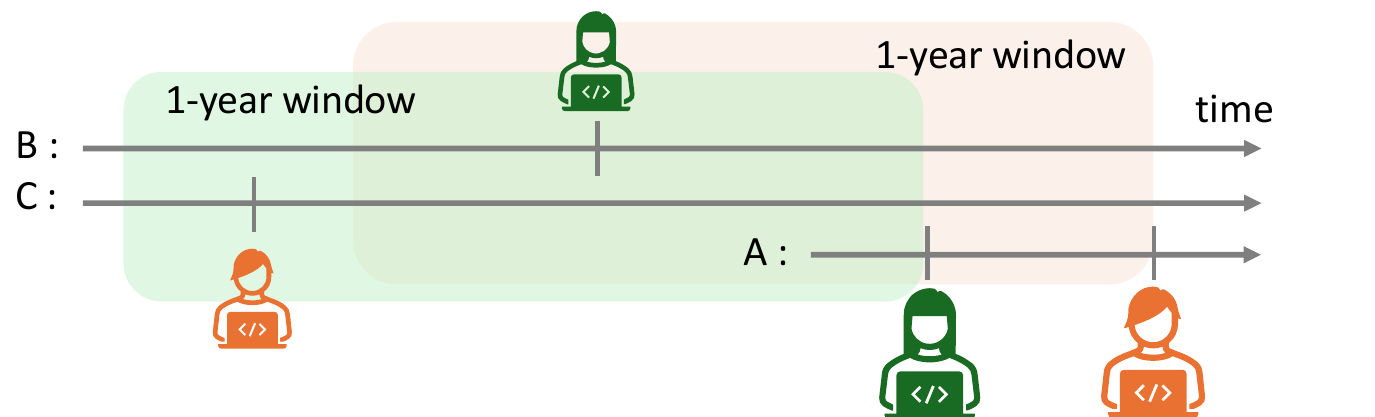}\vspace{-0.3cm}
\caption{Two core developers Green and Orange started contributing to a focal project $A$ on different dates. We record an edge from $A$ to $B$ (\ie $B$ could be a source of knowledge for $A$), because Green interacted with project $B$ in the previous year. We don't record an edge from $A$ to $C$, because Orange interacted with $C$ too far into the past.}
\label{fig:interaction-window}
\end{figure}

Each of the three actions we consider (making commits, participating in issue discussions, and starring) can be thought of as a person--to--project tie. 
Aggregating across all developers and projects in our sample leads to three person--to--project bipartite networks, one for each action type.
However, since our analysis is at the project level, we construct three directed project--to--project networks by way of projection, to represent possible flow of knowledge across projects. 
As discussed previously, we consider only actions initiated by core developers (\eg which other repositories they committed to or starred in the past).
In addition, we restrict the scope to actions in the recent past (\eg within the last year, we validate with alternative period length in the appendix), as we expect older actions to be less relevant.
This time window is developer-dependent.
That is, we record an edge $A \rightarrow B$ if and only if a core developer of project $A$ interacted with project $B$ within one year prior to their initial commit to project $A$ , as illustrated in Figure~\ref{fig:interaction-window}. 
The weight of each edge represents the number of core developers from project $A$ who engaged with project $B$ via a specific type of interaction.




As expected, the structural properties of the three networks support our reasoning for why commits, issues, and stars, respectively, capture ties of decreasing strength. 
Comparing the transitivity\footnote{In network science, \textit{transitivity} is a measure of the ratio of triangles to triads.} values of our three networks,
we observe approximately an order of magnitude ($10 \times$) difference between each pair of networks.
Specifically, the commit network displays the highest levels of transitivity, followed by the issue network, while the star network exhibits the lowest level; see Supplementary Materials for details.

\subsection{Network Measures}
\label{sec:network-variables}


Next, we compute variables related to the network position of each project to represent the \textit{amount} and \textit{diversity of knowledge} that may be accessible to its core developers through each network. 

First, we use the number of connected projects as a proxy for the \textbf{amount of knowledge accessible} to a given project.
More formally, for every project, we compute its out-degree centrality within each of the three networks, defined as the total sum of the weights of all its out-edges. 
This step resulted in computing the three variables listed in Table~\ref{table:variables} under \textit{\textbf{Degree Variables (original)}}.

Second, we use the Node2Vec graph embedding algorithm~\cite{grover2016node2vec}
to generate vector representations (embeddings) for each node in the network based on their topological position.
The process begins by generating sequences of random nodes from the graph using random walks.
Starting from a random node in the network, we sample the next node by randomly selecting a direct neighbor.
The likelihood of selecting each node is proportional to the edge weight between the candidate next node and the current node. 
This process continues until we have selected enough nodes to complete the walk (in our study, each walk comprises 20 nodes). 
We repeat this process to generate multiple walks (or sequences of nodes).
Subsequently, we use a skip-gram model~\cite{mikolov2013efficient}, commonly used to generate embeddings for words in natural language, on the generated walks to learn embeddings for each node. 

Next, for each project in the network, we examine all other projects that receive a directed edge from the focal project (denoted as set $P$). 
Our \textbf{knowledge diversity index}\footnote{
The index is undefined for projects with out-degree centrality below two. We exclude these from our regression when considering diversity metrics as independent variables.} is the average pairwise distance between any two projects in $P$:
$    D = \frac{\sum_{i,j \in P, i\ne j}{-sim(v_i, v_j)}}{|P|*(|P|-1)}, $
where $i$ and $j$ are projects in $P$, and $v_i$ and $v_j$ are their vector representations from the Node2Vec model. The distance between $v_i$ and $v_j$ is the negation of the cosine similarity of the two vectors.
This step resulted in computing the three variables listed in Table~\ref{table:variables} under \textit{\textbf{Diversity Variables (original)}}.

\subsection{Dimensionality Reduction}
\label{subsec: PCA}


Since they are designed to capture different concepts, we run PCA separately for the three degree variables and the three diversity variables.
As standard, we first log-transformed all input variables and scaled them to a mean of zero and a standard deviation of one.

\looseness=-1
The PCA resulted in three components for the degree variables and another three for the diversity variables, with the proportion of variance explained by each component listed in Table~\ref{table:PCA} (top).
Inspecting the table, we observe that the first two PCs cumulatively explain over 80\% of the variance in both groups, thus we decided to retain only two components each for the degree and diversity variables. 
That is, we will use these PCs (listed under \textit{\textbf{Degree Variables (post-PCA)}} and  \textit{\textbf{Diversity Variables (post-PCA)}} in Table~\ref{table:variables})  instead of the original variables in our regression models below.

To interpret the two PCs we turn to Table~\ref{table:PCA} (bottom), which lists the loadings of the principal components onto the original variables.\footnote{Principal components are linear combinations of the original variables, with the loadings indicating the contributing coefficients. For example, the first PC for the degree variables, $PC1_{\text{degree}}$, uses the coefficients in the first column in Table~\ref{table:PCA} (bottom): $PC1_{\text{degree}} = 0.60 * \textit{Deg}_\text{commit} + 0.61 * \textit{Deg}_\text{issue} + 0.52 * \textit{Deg}_\text{star}$.}
Inspecting the table we make the following observations.
First, the loadings onto the first principal component (PC1) are relatively consistent across networks for both the degree and the diversity variables. 
Thus, we interpret PC1 to represent the average degree (or diversity of ties) of the project across the three networks.
For example, drawing on our theoretical framework (Section~\ref{sec:theory}), a project for which $PC1_{\text{degree}}$ is high is expected to have many ties (on average, across the three networks), \ie its core developers should have more sources from which to draw inspiration.
Similarly, the ties of a project for which $PC1_{\text{diversity}}$ is high are expected to span more of the knowledge embedding space we used to estimate tie diversity, \ie its core developers could have access to more varied sources to draw inspiration from.

\begin{table}[t]
    \setlength{\tabcolsep}{2.5pt}
    \renewcommand{\arraystretch}{1}
    \caption{\textit{Top}: Proportion of variance explained by each principal component. \textit{Bottom}: Loadings of the principal components onto the original degree and diversity variables.} 
    \label{table:PCA} 
    \centering \small \vspace{-0.2cm}
  \begin{tabular}{L{3cm} r r r r r r r} 
  \toprule
   & \multicolumn{3}{c}{\textbf{Out-deg.\ centrality}} & & \multicolumn{3}{c}{\textbf{Diversity index}} \\ 
   \cmidrule(lr){2-4} \cmidrule(lr){6-8}
 & \textbf{PC1} & \textbf{PC2} & \textbf{PC3} & & \textbf{PC1} & \textbf{PC2} & \textbf{PC3} \\
  Variance Explained & 0.64 & 0.22 & 0.14 & & 0.51 & 0.29 & 0.19 \\ 
  Cumulative Variance & 0.64 & 0.86 & 1.00 & & 0.51 & 0.81 & 1.00 \\ 
  \midrule
$\text{D}_{\text{commit}}$ &  0.60 & $-$0.45  & 0.67 & & 0.63 & $-$0.36 & 0.69 \\ 
$\text{D}_{\text{issue}}$ & 0.61 &  $-$0.28  & $-$0.74 & & 0.65 & $-$0.24 & $-$0.72  \\  
$\text{D}_{\text{star}}$ & 0.52 &  0.85 & 0.11 & & 0.43 & 0.90 & 0.08\\ 
  \bottomrule
  \end{tabular} 
  \end{table}

Second, the loadings onto the second component (PC2) show a descending trend across networks, again similarly for both sets of variables. 
Notably, the highest loading on the degree (or diversity) comes from the star network, followed by the issue network, while the lowest comes from the commit network.
Thus, we interpret PC2 to represent the strength of network ties in the degree (or diversity of ties) metric, \ie the \textbf{strength of weak ties}. 
A project for which $PC2_{\text{degree}}$ is high is expected to get more of its connectivity through the star network.
Analogously, relatively more of the diversity of knowledge accessible to core developers in a project for which $PC2_{\text{diversity}}$ is high can be attributed to the star network (weak ties) compared to the commit or issue networks (stronger ties).

It may seem counterintuitive to reason about PC1 and PC2 jointly, especially in a regression modeling framework. 
What does it mean to vary PC2 from low to high, for instance, while holding PC1 fixed? 
Can PC1 and PC2 even vary independently? 
They can in theory, because they are orthogonal by construction, but are the combinations of low-high values of PC1 and PC2 observable in the real-world data? 
After all, regression is an interpolation mechanism.
We illustrate this space of interpretations of PC1 and PC2 for the diversity variables with the artificial examples in Figure~\ref{figure:pca-ties} (see Supplementary for real-world examples in each quadrant).
In the top left quadrant (PC1 and PC2 both high), the project core developers can tap into a diverse knowledge space via either strong or weak ties. In the top right (PC1 high and PC2 low), the overall diversity of information accessible is just as high, but it's the strong ties that are the source of it. In the bottom left (PC1 low and PC2 high), the overall diversity of information available is lower, but whatever diversity there is, it can be attributed mostly to the weak ties. Finally, in the bottom right (PC1 and PC2 both low), weak ties contribute little to an otherwise low diversity of information.

Going back to our hypotheses (Section~\ref{sec:theory}), we expect the left quadrants, where weak ties are strong, to be most associated with innovativeness. As an anecdote, consider the example in Figure~\ref{fig:geonotebook} on the first page. The GeoNotebook project is a Jupyter notebook-based environment for interactive visualization and analysis of geographic data. Interestingly, GeoNotebook's core developers have previously starred (without otherwise contributing to) many GitHub projects spanning a variety of seemingly related topics, including a collection of free APIs, example data science notebooks, reproducible workflows for Jupyter notebooks, visualization tools, time series analysis tools, geographic data analysis tools, and infrastructure tools for interactive packet manipulation. One can expect that this diverse space of ideas provided some inspiration, knowingly or unknowingly, for the design and implementation of GeoNotebook. Our analysis below tests to what extent there is evidence supporting this mechanism at scale, beyond this anecdote.

\begin{figure}[t]
\centering
\includegraphics[width=0.9\columnwidth, clip=true, trim=0 175 860 0]{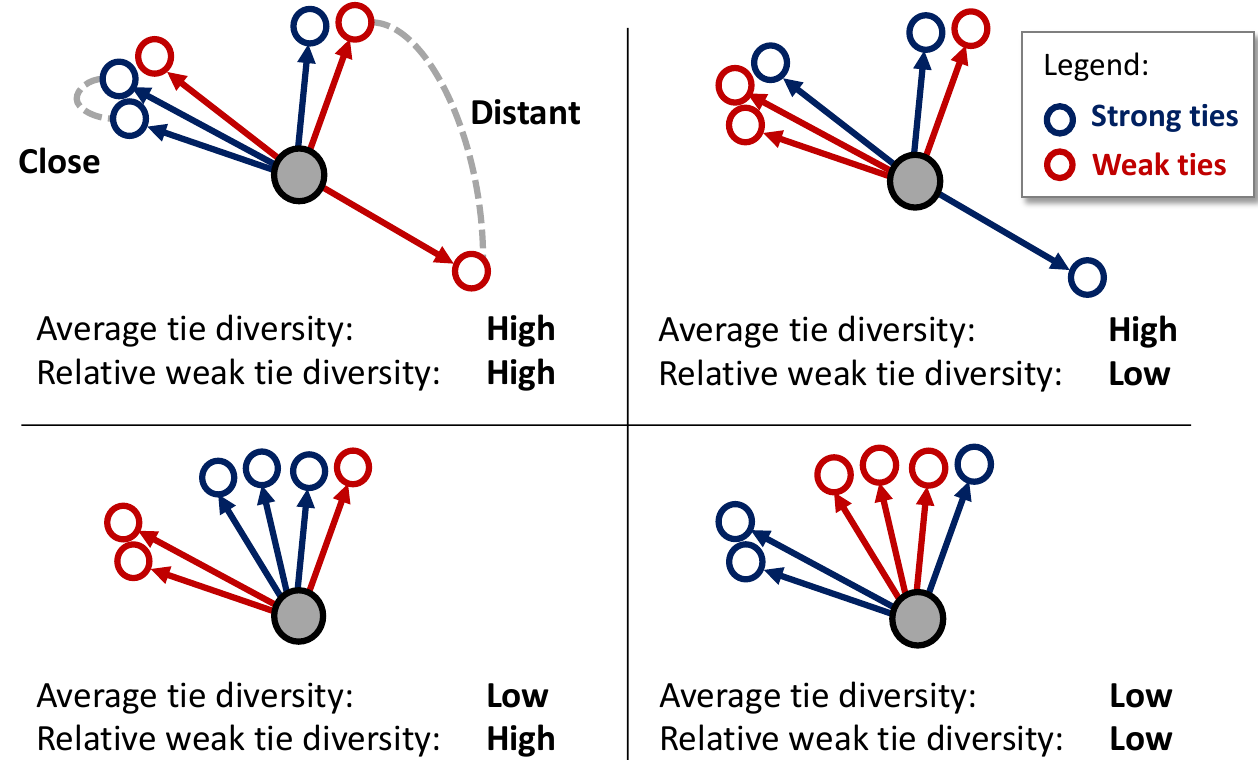}
\caption{Illustration of the four possible combinations of low-high values for $PC1_{\text{diversity}}$ (corresponding to average tie diversity) and $PC2_{\text{diversity}}$ (corresponding to the relative diversity attributable to weak ties).}
\label{figure:pca-ties}
\end{figure}

\begin{table}[t]
\renewcommand{\arraystretch}{0.9}
\setlength{\tabcolsep}{2pt}
\centering
\small
\caption{The definitions of the variables in our models.}
\label{table:variables}
\vspace{-0.2cm}
\begin{tabular}[t]{P{2.1cm}p{6.5cm}}
\toprule
  \multicolumn{2}{l}{\textit{\textbf{Outcome Variables}}} \\ 
\textit{Innovativeness} & The extent to which a project imports packages that were typically not combined in the past, computed based on all the packages imported until the end of 2021. \\

    \multicolumn{2}{l}{\textit{\textbf{Degree Variables (original)}}} \\ 
$\text{\textit{Deg}}_{\text{Commit}}$ & The sum of edge weights from the focal project to the others in the commit network. \\ 
$\text{\textit{Deg}}_{\text{Issue}}$ & The sum of edge weights from the focal project to the others in the issue network. \\ 
$\text{\textit{Deg}}_{\text{Star}}$ & The sum of edge weights from the focal project to the others in the star network. \\ 

    \multicolumn{2}{l}{\textit{\textbf{Degree Variables (post-PCA)}}} \\ 
$\textit{Deg}_{\text{Ave}}$ (\hr{hyp1}) & The first principal component resulted from the PCA analysis over three original degree variables, representing average network degree across the three networks. \\ 
$\textit{Deg}_{\text{Weakness}}$ & The second principal component resulted from the PCA analysis over three original degree variables, representing the degree from weak ties. \\


    \multicolumn{2}{l}{\textit{\textbf{Diversity Variables (original)}}} \\ 
$\text{\textit{Div}}_{\text{Commit}}$ & The diversity of projects to which the focal project has a directed edge in the commit network. \\ 
$\text{\textit{Div}}_{\text{Issue}}$ & The diversity of projects to which the focal project has a directed edge in the issue network. \\ 
$\text{\textit{Div}}_{\text{Star}}$ & The diversity of projects to which the focal project has a directed edge in the star network. \\ 

    \multicolumn{2}{l}{\textit{\textbf{Diversity Variables (post-PCA)}}} \\ 
$\textit{Div}_{\text{Ave}}$ (\hr{hyp2}) & The first principal component resulted from the PCA analysis over three original diversity variables, representing average diversity of connections across the three networks. \\ 
$\textit{Div}_{\text{Weakness}}$ (\hr{hyp3}) & The second principal component resulted from the PCA analysis over three original diversity variables, representing the diversity from weak ties. \\

  \multicolumn{2}{l}{\textit{\textbf{Control Variables}}} \\ 
$\textit{Year}_{\text{creation}}$ & A fixed effect variable indicating the year in which the project received its first commit. \\
\textit{Org\_owned} & A binary variable indicating whether the project was owned by an organizational account or not. \\
$N_{\text{Owner\_Stars}}$ & The total number of stars that the project owner received on all other repositories they owned before the creation of the focal project. \\
$N_{\text{Core\_Devs}}$  & The total number of core developers in the project, as computed before the end of 2021. \\
$N_{\text{Packages}}$  & The number of software packages that the project imported before the end of 2021. \\
\bottomrule
\end{tabular}
\vspace{-0.5cm}
\end{table}

\subsection{Project Innovativeness Measure}



\looseness=-1

For each project, we extract the sequence of packages imported\footnote{We ignore built-in Python libraries like~\citet{fang2024novelty} did, and consider only packages published in the PyPI registry, using the explicit links back to GitHub recorded there.} (precomputed in World of Code) and use a skip-gram model to generate positional embeddings for each package based on these dependency relationships. 
These embeddings place the package in an optimal location relative to all other packages imported in the same project, based on their co-import relationships.
Unlike the random walks used in the Node2Vec model to generate embeddings for the three interaction networks, there is no inherent order in the sequence of packages imported within the same project. 
Therefore, we set the window size in the skip-gram model to be sufficiently large, enabling it to consider all packages in the same sequence as its context when generating embeddings.
Similar approaches to compute embeddings of software packages have been used before in software engineering research~\cite{dey2021representation, fang2023matching}.

To estimate the atypicality of a combination of two packages, we compute the negation of their embeddings' cosine similarity. 
Then, our \textbf{measure of project innovativeness} is the average pairwise atypicality of all pairs of packages imported by a project.


\mysec{Validation}
\label{validation awesome-atypicality}
Although measuring project novelty in terms of the atypicality of package combinations therein has solid theoretical foundation (Section~\ref{sec:theory}) in addition to application in past empirical studies of open-source software~\cite{fang2024novelty}, this likely remains the most controversial part of our methodology.
Therefore, we perform two validation checks, comparing to an existing dataset of ``awesome'' projects and to the previous measure by \citet{fang2024novelty}.

``Awesome'' lists are community-curated lists of software. 
Many exist for different programming languages, platforms, application domains, etc. 
One can propose new entries via a pull request (PR), and typically some minimum amount of support (+1s) is required for the PR to be merged.
Being innovative isn't formally a requirement for being ``awesome,'' although as part of the same PRs submitters usually argue that the proposed entries are novel (\eg PRs to the Python list contain a section titled ``What's the difference between this Python project and similar ones?'')

\begingroup

\setlength{\intextsep}{6pt}     
\setlength{\columnsep}{12pt}    

\begin{wrapfigure}{R}{0.5\columnwidth}
  \begin{center}
    \includegraphics[width=0.49\columnwidth, clip=true, trim= 0 0 0 0]{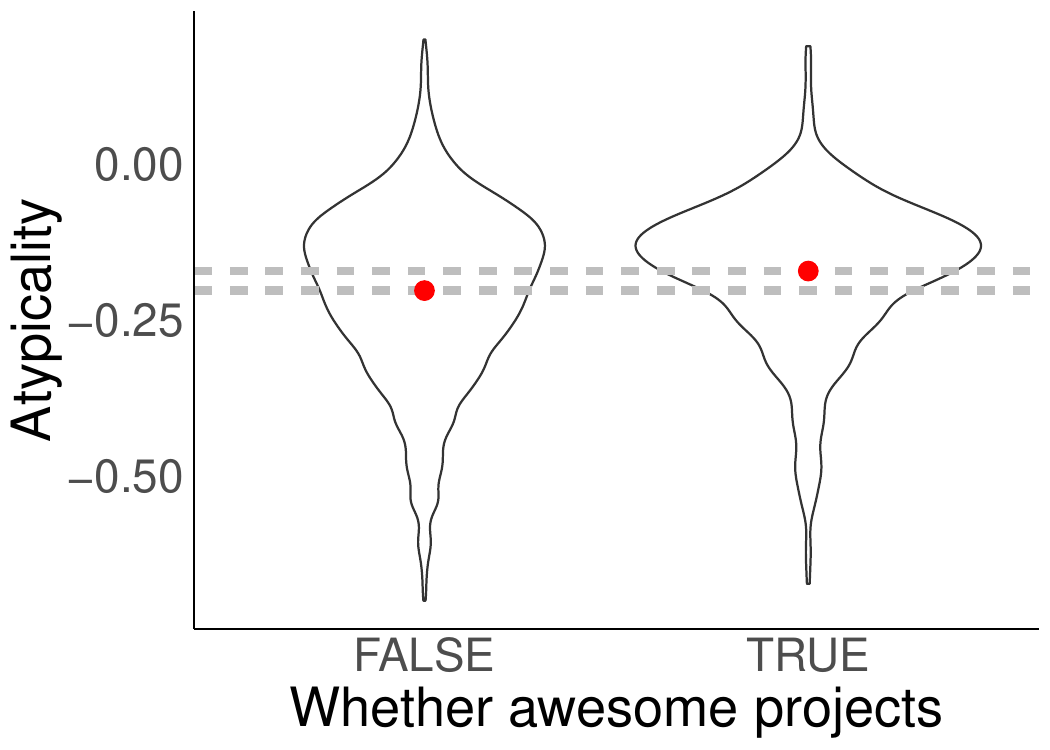}
  \end{center}
  \caption{\small ``Awesome'' projects have statistically significantly higher atypicality scores than the rest of our sample. The red dots represent the distribution means.}
  \label{figure:awesome_list}
\end{wrapfigure}

We scraped software mentions from a popular aggregator of ``awesome'' lists~\citep{awesome}
(350k+ GitHub stars at the time of writing), and identified 760 Python projects matching our sampling criteria (\S\ref{subsec:dataset}), and for which we could compute atypicality scores (\eg having at least two import-based dependencies we could identify), corresponding to about 2\% of our sample.
Comparing the atypicality scores between the ``awesome'' and remaining projects in our sample (Figure~\ref{figure:awesome_list}), we observe a statistically significant difference in means (two-sample independent t-test $p < 8.8e^{-13}$), with a non-trival effect size: a project with a median atypicality score in the non-selected group will only be among the bottom 35\% ordered by atypicality score among the ``awesome'' projects. 
Further validation suggests this effect is not confounded by variables such as project size (Supplementary).
Therefore, we conclude that \textbf{our innovativeness measure reflects to some extent the developers' perception of ``awesome'' projects, which usually includes a dimension of novelty.}

In addition, we compute the correlation between the atypicality scores obtained using our skip-gram model and those generated using the original measure by \citet{fang2024novelty}. Results show a moderate-to-strong~\cite{schober2018correlation} linear (Pearson) correlation of 0.65.
Therefore, we conclude that \textbf{our innovativeness measure correlates to that of prior work to a large degree.}
Furthermore, we test the robustness of our conclusions to variations in the innovativeness measure, and find that the estimated regression coefficients for the \textit{degree} and \textit{diversity} variables behave consistently with both measures.
Thus, we only present the results obtained with our proposed measure below, but include the corresponding regression results with the original measure by \citet{fang2024novelty} in Supplementary.

\endgroup

\subsection{Regression Modeling Considerations}




To test our three hypotheses, we use fixed-effects linear regression (one data point per project), where the response variable is our measure of \textit{innovativeness}, and our main explanatory variables are the degree and diversity principal components discussed above:
$\textit{Deg}_{\text{Ave}}$ captures \hr{hyp1} (volume of interactions the core developers had with other projects in the previous year), 
$\textit{Div}_{\text{Ave}}$ captures \hr{hyp2} (diversity of information accessible),
and $\textit{Div}_{\text{Weakness}}$ captures \hr{hyp3} (strength of weak ties).
The remaining principal component, $\textit{Deg}_{\text{Weakness}}$, does not directly map to one of our hypotheses, but can help distinguish whether it's the volume of weak ties, or their strength, that explains more variance in our outcome measure.

We also control for several important confounding variables (Table~\ref{table:variables}): the repository owner's social standing ($N_{\text{Owner\_Stars}}$), the project team size ($N_{\text{Core\_Devs}}$), the complexity of the codebase in terms of number of packages imported ($N_{\text{Packages}}$), whether the project is owned by an organizational account (\textit{Org\_owned}), and the project age ($\textit{Year}_{\text{creation}}$, modeled as a categorical variable).
For model fit and diagnostics we followed standard practice, \eg we first log-transformed the variables with skewed distributions to reduce heteroskedasticity~\cite{kaufman2013heteroskedasticity}; see replication package.


\section{Results}
\label{sec:results}

Next we present the main results from the regression analysis formally testing our hypotheses, summarized in Table~\ref{table: regression analysis}, as well as a series of robustness checks.


\begin{table}[t] 
\centering \small
\renewcommand{\arraystretch}{0.6}
\setlength{\tabcolsep}{2pt}
  \caption{Summary of regression analysis of factors associated with differences in project innovativeness.} 
  \label{table: regression analysis} 
\vspace{-0.2cm}
\begin{tabular}{P{3cm}cccc} 
\toprule
& Model I & Model II & Model III & Model IV\\ 
\midrule
  \textit{\textbf{Variables of interest}} & & & & \\
$\textit{Deg}_{ave}$ (\hr{hyp1}) & 0.001$^{***}$ & $-$0.001 &  & $-$0.002$^{**}$ \\ 
  & (0.0002) & (0.0006) &  & (0.0006) \\ 
  & & & & \\ 
$\textit{Deg}_{weakness}$ & $-$0.0003 & $-$0.001 &  & $-$0.005$^{***}$ \\ 
  & (0.0002) & (0.0007) &  & (0.0008) \\ 
  & & & & \\ 
 $\textit{Div}_{ave}$ (\hr{hyp2}) &  &  & 0.006$^{***}$ & 0.007$^{***}$ \\ 
  &  &  & (0.0006) & (0.0006) \\ 
  & & & & \\ 
  $\textit{Div}_{weakness}$ (\hr{hyp3}) &  &  & 0.005$^{***}$ & 0.007$^{***}$ \\ 
  &  &  & (0.0008) & (0.0008) \\ 
  \textit{\textbf{Controls}} & & & & \\
 \textit{Org\_owned} & 0.027$^{***}$ & 0.017$^{***}$ & 0.018$^{***}$ & 0.018$^{***}$ \\ 
  & (0.0006) & (0.0016) & (0.0016) & (0.0016) \\ 
  & & & & \\ 
 $N_{\text{Owner\_Star}}$ (log)  & 0.002$^{***}$ & 0.001$^{*}$ & 0.001 & 0.001$^{*}$ \\ 
  & (0.0002) & (0.0004) & (0.0004) & (0.0004) \\ 
  & & & & \\ 
 $N_{\text{Core\_Dev}}$ (log) & 0.014$^{***}$ & 0.018$^{***}$ & 0.017$^{***}$ & 0.018$^{***}$ \\ 
  & (0.0006) & (0.0016) & (0.0015) & (0.0016) \\ 
  & & & & \\ 
 $N_{\text{Packages}}$ (log) & 0.042$^{***}$ & 0.023$^{***}$ & 0.023$^{***}$ & 0.024$^{***}$ \\ 
  & (0.0003) & (0.0011) & (0.0011) & (0.0011) \\ 
 & & & & \\ 
 \textit{\textbf{Fixed effect}} & & & & \\
$\text{Year}_{\text{creation}}$ & \checkmark & \checkmark & \checkmark & \checkmark \\ 
\midrule
Observations & 589,999 & 37,451 & 37,451 & 37,451 \\ 
Adjusted R$^{2}$ & 0.054 & 0.059 & 0.063 & 0.064 \\ 
\bottomrule
&   \multicolumn{4}{r}{$^{*}$p$<$0.05; $^{**}$p$<$0.01; $^{***}$p$<$0.001} \\ 
\end{tabular} 
\vspace{-0.5cm}
\end{table}







\mysec{Main Regression Results}
Model I (all projects) and Model II (projects with a minimum out-degree centrality of two across all three networks) shed light on the impact of degree variables. Model~I reveals a significant positive effect from average degree centrality, whereas Model II shows a negative effect. Nevertheless, effect sizes for both models are relatively small. For instance, in Model I, transitioning a project's $\textit{Deg}_{ave}$ from the 25th percentile (-1.09) to the 75th (0.53) results in a mere 0.002 increase in project innovativeness.
Considering that the 25th percentile of a project's innovativeness score is -0.33, and the 75th is -0.12 in our sample, the change in project innovativeness due to variations in average degree centrality is practically negligible. 
Therefore, \hr{hyp1} is not supported.

\vspace{-1mm}
\begin{tcolorbox}[colback=white, colframe=black, boxsep=2pt, left=2pt, right=2pt, top=2pt, bottom=2pt]
We do not find supporting evidence that the more interactions developers have with other developers and projects, the more innovative their projects are (\hr{hyp1}).
\end{tcolorbox}

In Model III, we observe a positive association between the average diversity ($\textit{Div}_{ave}$) of ties originating from the focal project and project innovativeness. Moreover, this effect is notably stronger -- nearly seven times greater -- compared to that of average degree in Models I and II. 
Transitioning a project from the 25th percentile (-0.89) to the 75th (0.86) in terms of $\textit{Div}_{ave}$ corresponds to an increase of approximately 0.012 in project innovativeness, representing roughly a 4\% change in the distribution (\eg a shift from the median project in terms of novelty to the 54th percentile).
We conclude that \hr{hyp2} is supported.

\vspace{-2mm}
\begin{tcolorbox}[colback=white, colframe=black, boxsep=2pt, left=2pt, right=2pt, top=2pt, bottom=2pt]
On average, the greater the informational diversity of developers' past interactions, the more innovative their projects are (\hr{hyp2}).
\end{tcolorbox}

Additionally, we observe a significant effect for the $\textit{Div}_{weakness}$ variable. This suggests that, when controlling for the average diversity across the three networks, higher diversity in the ``weak ties network'' (\ie star network), or stronger weak ties, corresponds to increased project innovativeness. 
In practical terms, the effect size for the $\textit{Div}_{weakness}$ variable is comparable, albeit somewhat smaller, to that of the $\textit{Div}_{ave}$ variable above.
We interpret this as supporting evidence for \hr{hyp3}.

\vspace{-1mm}
\begin{tcolorbox}[colback=white, colframe=black, boxsep=2pt, left=2pt, right=2pt, top=2pt, bottom=2pt]
On average, the more the informational diversity of developers' past interactions is due to weak ties, the more innovative their projects are (\hr{hyp3}).
\end{tcolorbox}

When incorporating both degree and diversity variables in the same regression model, as in Model IV, the positive effects of $\textit{Div}_{ave}$ and $\textit{Div}_{weakness}$ remain significant. This suggests a consistent effect stemming from tie diversity and strength of weak ties. Nevertheless, we also note a significant negative effect for $\textit{Deg}_{ave}$ and $\textit{Deg}_{weakness}$ after adjusting for project diversity.
These negative effects suggest that, while holding tie diversity constant, higher out-degree centrality and more ties in the ``weak network'' are associated with decreased project innovativeness.
This observation could be attributed to the increased constraints within a local community associated with higher out-degree, as discussed by Burt~\cite{burt2000network}. 
Such constraints can lead to a reduced inclination for combining packages of diverse functions within the project, consequently contributing to the reduced novelty of the project.
More research is needed to further test these effects.

\begin{figure*}[t]
    \centering

    \begin{subfigure}[b]{0.31\textwidth}
        \centering
        \includegraphics[width=\textwidth]{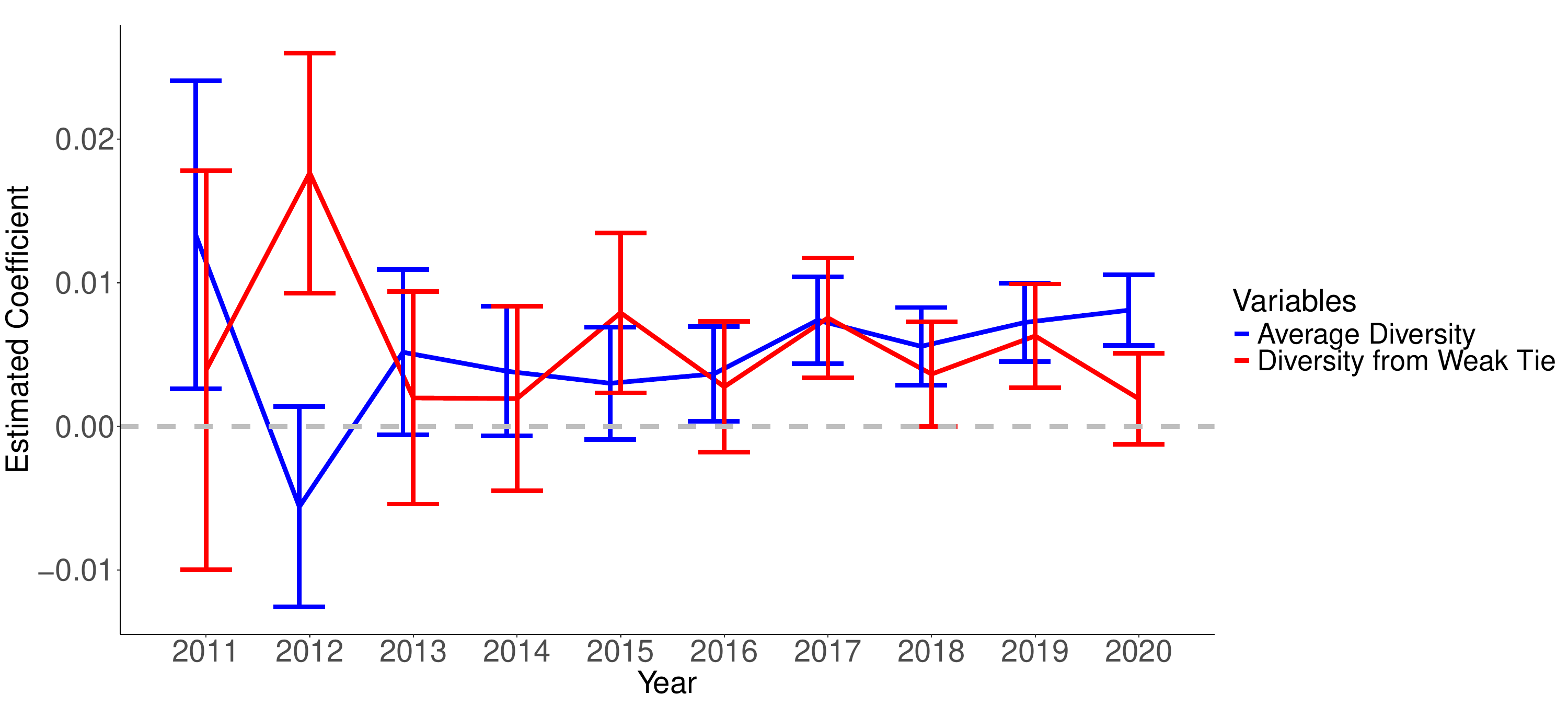}
        \caption{Effects of knowledge diversity across years.}
        \label{fig: diversity year cohort}
    \end{subfigure}
    \quad
    \begin{subfigure}[b]{0.31\textwidth}
        \centering
        \includegraphics[width=\textwidth]{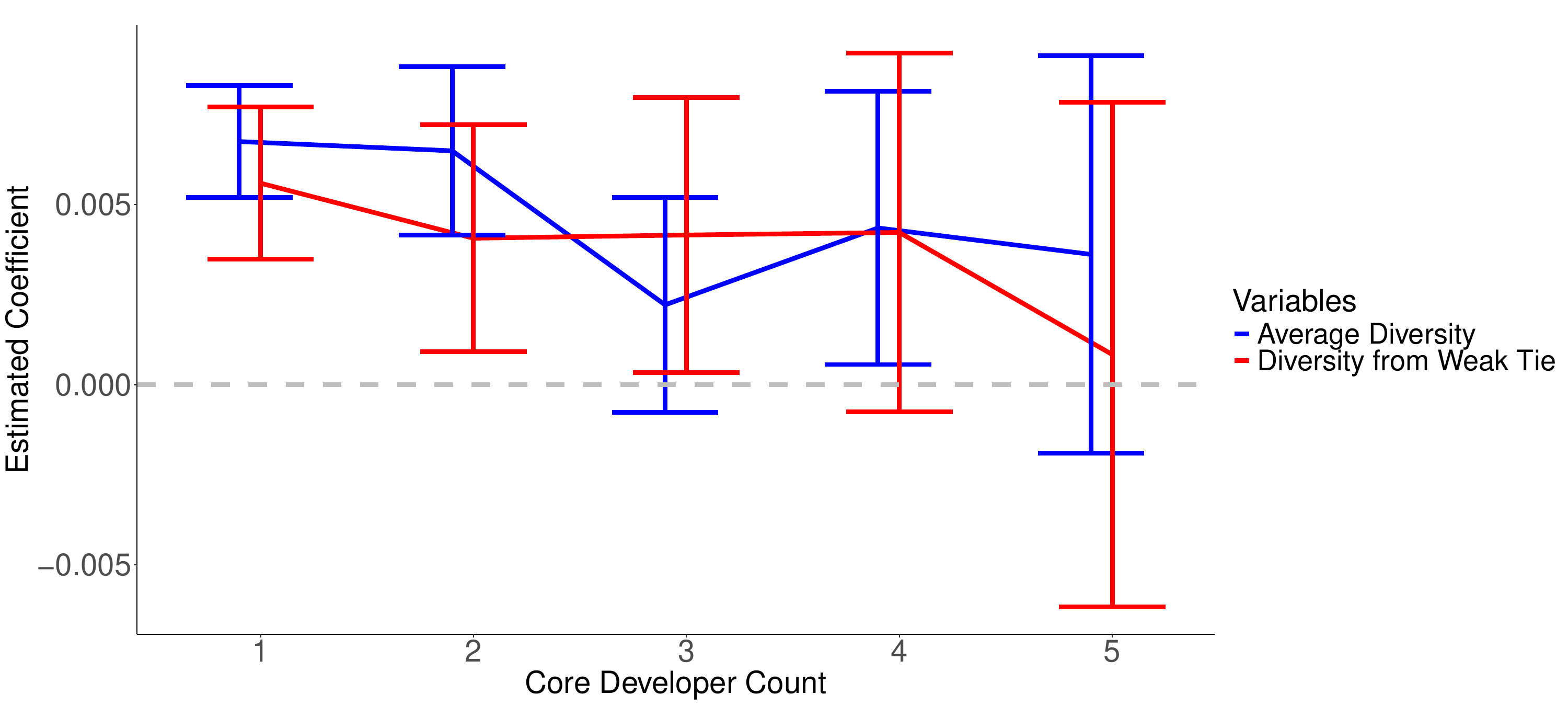}
        \caption{Effects of knowledge diversity on different team-size projects.}
        \label{fig:diversity core cohort}
    \end{subfigure}
    \quad
    \begin{subfigure}[b]{0.31\textwidth}
        \centering
        \includegraphics[width=\textwidth]{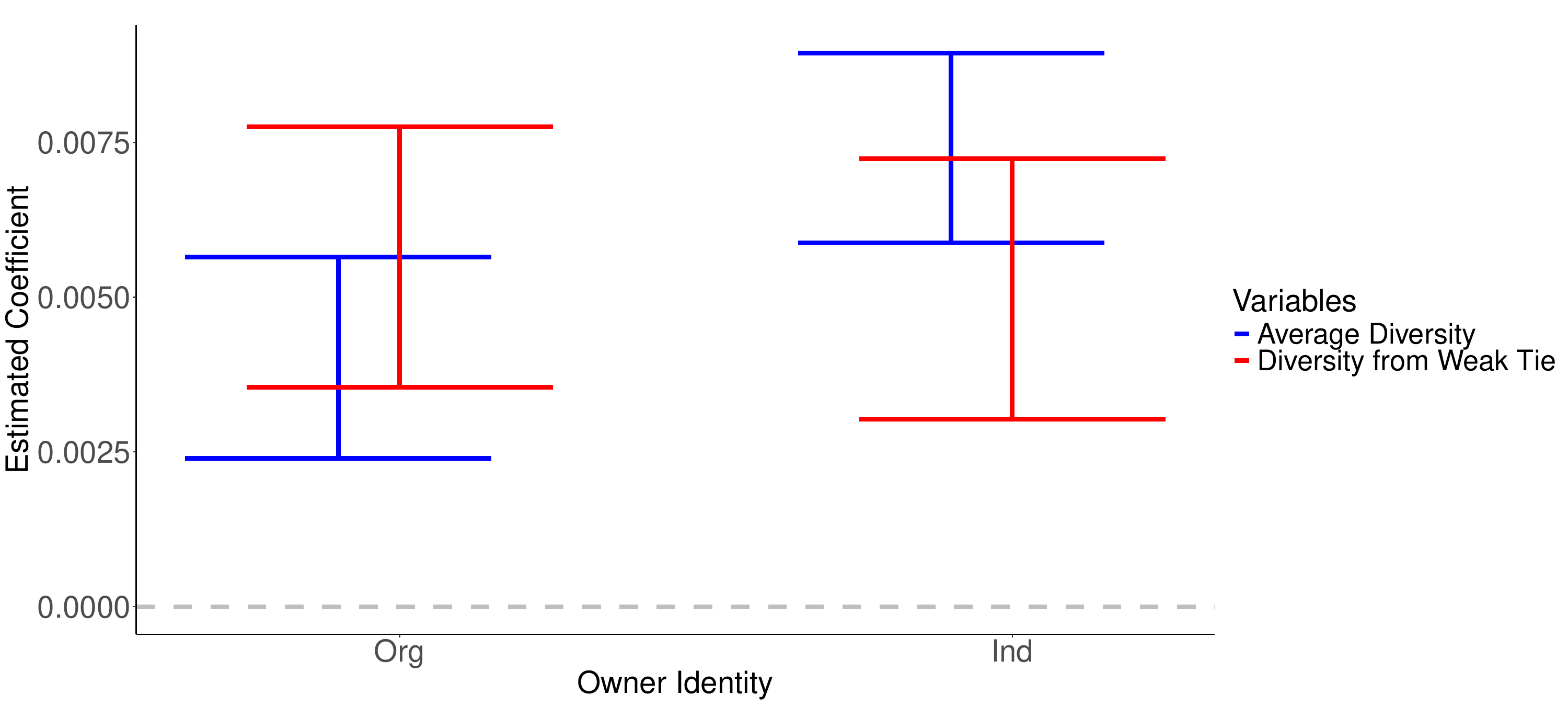}
        \caption{Effects of knowledge diversity on individual vs org projects.}
        \label{fig:diversity org cohort}
    \end{subfigure}
    
    \caption{The effects of the two diversity of knowledge variables ($\textit{Div}_{ave}$ testing \hr{hyp2} and $\textit{Div}_{weakness}$ testing \hr{hyp3}) are generally robust. Error bars denote 95\% confidence intervals.}

\end{figure*}



\mysec{Robustness Checks}
In addition to the regressions above, we test how robust our conclusions about the effects of $\textit{Div}_{ave}$ and $\textit{Div}_{weakness}$ are to differences in operational decisions.


\mysubsec{Does Time of Measurement Matter? ``No.''}
We segmented our dataset by project inception year and conducted regression analyses with the specification of Model III in Table~\ref{table: regression analysis} on projects initiated in each respective year. 
Figure~\ref{fig: diversity year cohort} shows the resulting estimated coefficients for the two diversity variables. 
As illustrated in the graph, the positive relationship between the average diversity of knowledge access across the three networks (represented by the blue line) remains consistent over the years, with the average estimated coefficient being statistically significantly above zero in most years. 
The estimated coefficient for $\textit{Div}_{ave}$ appears to also increase in recent years, though this difference is not statistically significant.

The positive relationship between diversity from weak ties and project innovativeness is generally consistent (the red line in Figure~\ref{fig: diversity year cohort}). 
The estimated coefficient is above zero in all ten years observed, and the statistically insignificant coefficient (\ie the lower bound of the 95\% confidence interval is below zero) is likely attributable to the limited number of projects in each annual cohort (\ie fewer than 4,000 projects in each year before 2016).



\mysubsec{Does Project Size Matter? ``No.''}
Similarly, we split projects into subsets based on the number of core developers and assessed the impact of the diversity variables in each cohort. 
As shown in Figure~\ref{fig:diversity core cohort}, the effect of $\textit{Div}_{ave}$ remains consistent across projects of varying team sizes, confirming the robustness of our findings. It is noteworthy that the effect of $\textit{Div}_{weakness}$ becomes insignificant for projects with four or more core developers. This may be attributed to the limited number of project samples within these categories.





\mysubsec{Does Project Ownership Matter? ``No.''}
Lastly, we re-estimated the effects of the diversity variables separately for projects owned by organizational and individual accounts. 
Our analysis (Figure~\ref{fig:diversity org cohort}) shows that both \textit{average diversity across networks} and \textit{diversity from weak ties} have significantly positive effects for both categories. Furthermore, the difference in effect size between organization-owned and individual-owned projects is not statistically significant.


\vspace{-2mm}
\begin{tcolorbox}[colback=white, colframe=black, boxsep=2pt, left=2pt, right=2pt, top=2pt, bottom=2pt]
In conclusion, the positive associations between average knowledge diversity and project innovativeness, as well as between diversity from weak ties and innovativeness, are robust.
\end{tcolorbox}

\section{Discussion}
\label{sec: discussion}




Next we discuss our results in the broader context of the literature and highlight budding directions for future research and practice.

\mysec{Summary of Main Results}
Established social science theory describes the mechanisms through which social network weak ties can be particularly ``strong'' -- they can bridge disjoint parts of the network, providing access to diverse information, which in turn can be used towards better outcomes.
Our study provides the first empirical link (that we know of) between this theoretical mechanism and an important software engineering outcome -- the emergence of innovation in a software project.
We hypothesized that the many user-to-user and user-to-artifact interactions possible on the GitHub platform, ranging from high-effort ones like making changes to a codebase to relatively trivial ones like starring a repository, result in network ties of varying strength, along which knowledge can flow. And, in particular, that weak ties are instrumental for this knowledge flow, as the GeoNotebook anecdote in the Introduction (Figure~\ref{fig:geonotebook}) would suggest.

Using a sophisticated methodology to reconstruct interaction networks, estimate tie strength, compute network informational diversity, and estimate project innovativeness, coupled with a robust statistical analysis, we found clear evidence (albeit only correlational) supporting two of our three hypotheses: 
on average, it's not the amount (\hr{hyp1}), but rather both the diversity of information available to a focal project's core developers (\hr{hyp2}), as well as the extent to which this information diversity comes from weak ties (\hr{hyp3}), that explain some of the variance in how novel (atypical) the combination of software packages imported in a focal project is. 

\mysec{Effect Sizes}
Some readers may question whether our main variables explain \textit{enough} variance, but small effect sizes are exactly what we should expect when studying innovation in complex socio-technical systems. 
Software development operates in an extraordinarily high-dimensional space where countless factors influence outcomes, making any single mechanism necessarily modest in isolation. 
The theoretical mechanism of weak ties operates through subtle informational advantages that accumulate over time rather than dramatic singular influences, and these effects must compete against an already information-rich environment of documentation, tutorials, and online resources. 
Moreover, as discussed in the beginning of our paper, the weak ties theory has decades of robust empirical support across diverse domains, consistently showing modest but meaningful effects that compound over time. 
We're far from the first to discover this effect -- at best, we're the first to offer some supporting evidence from the open-source software development context.
In addition, most software projects engage in predictable recombination of existing packages, with genuinely novel combinations occurring primarily in the tail of the distribution where even small probability shifts can have outsized impacts. 
The consistency of effects across multiple strata, coupled with statistical significance despite the noisy environment, actually strengthens confidence in the underlying mechanism -- large effect sizes in network research are often suspicious because they suggest implausibly crude influences that overwhelm all other factors, which rarely occurs in real complex systems.
Next, we discuss the implications of our results.




\mysec{Platform Design}
One way to interpret the strength-of-weak-ties theory in our context is that lurking on the GitHub platform (\ie we see starring repositories as lurking rather than actively contributing) has quantifiable benefits.
If indeed people draw inspiration from things they've starred and apply those ideas to something they will build next, as the theory would suggest and our quantitative results support, platform designers should consider facilitating this process more. 
For example, GitHub has a way of highlighting ``what the community is most excited about today'' on its Trending page. 
It could be beneficial if this algorithm that determines what is trending (allegedly based on the star growth rate) also took into account and increased the overall informational diversity of the highlighted repositories (as computed, \eg using an embedding-based approach like ours).
Better still, the recommendations could be personalized, \ie the algorithm could highlight repositories that enhance information diversity \textit{for a given user}.

Our results also raise questions about the GitHub culture of (implicitly) rewarding code contributions the most, since one's commit history tends to be highly visible in the platform. 
Much has been written about how other developers~\cite{marlow2013impression} and even recruiters~\cite{capiluppi2012assessing} use such signals. 
In contrast, we are finding evidence that well-informed but not necessarily highly active developers may also be experts at their craft, at least insofar as the novelty of their output.
It's worth thinking about how to feature (and reward) such lurking more prominently in the platform design.

It's also worth thinking about what tracking and giving credit to ideas, not just code artifacts, might look like in this domain. 
In scientific research, authors are expected to cite prior work that influenced their thinking. 
In turn, these citations form networks encoding invaluable information about scientific progress and scientific collaboration~\cite{fortunato2018science}. 
What might such a system look like in software development? 
Anecdotally, it seems like developers already use many ad-hoc ways to credit ideas, \eg including links in source code comments~\cite{hata20199} and even referring to research papers in repository README files~\cite{wattanakriengkrai2022github} or source code~\cite{inokuchi2019academia}. 
But we still don't understand how common the practice is, how complete the mentions are, and what the supply chains of ideas look like.
Nor do we have ways of systematically tracking ideas as part of the platform design, which is worth exploring.

\mysec{Theory}
More broadly, there are many opportunities for developers for relatively low-effort interaction with each other and with new technology besides the one we measured here (starring). 
These include using social media to stay current~\cite{singer2014software, fang2022damn, wyrich2024beyond}, listening to podcasts~\cite{engzell2023podcasts}, creating or consuming video content online~\cite{chattopadhyay2021reel}, and participating in developer conferences~\cite{truong2022unsolvable} or trainings.
Since our study is not exploratory (bottom-up), but rather tests hypotheses drawn from a much more general, widely-validated theory (top-down), it's reasonable to expect, given the theory, that the same ``strong weak ties'' mechanism could also apply in these scenarios.
The famous ``water-cooler effect,'' that advocates of in-person work environments often quote, is another example of the same mechanism -- casual encounters, typically with weak ties (work acquaintances rather than close colleagues), can help exchange new ideas that spark creativity~\cite{brucks2022virtual}. 
We need more research to empirically investigate these and other types of weak ties and their innovation-enabling potential, as it would be important not only to collect more evidence in support of the mechanism, but also to rank the different interactions in terms of their effectiveness.
For example, many technology companies support their engineers regularly attending conferences and these experiences are likely beneficial.
Could we quantify these benefits both in general, and relative to, say, consuming technical social media content?

Separately, our work aligns well with Baltes and Diehl’s~\cite{baltes2018towards} software development expertise theory. For example, our lurking can be considered as an example of \textit{behavior} that contributes to a developer's knowledge base, or as an example of \textit{continuous learning}; 
similarly, the GitHub platform is the \textit{work context} that influences expertise development, by facilitating exposure to diverse projects.
But perhaps more interestingly, while the theory emphasizes \textit{deliberate practice} for expertise development, our work offers a complementary perspective and an opportunity for theory refinement: casual observation of diverse projects may contribute to innovative thinking in ways that focused practice alone might not.


\mysec{Exploration vs Exploitation}
Research has shown that developers are more likely to join projects that are technically more familiar to them~\cite{dey2021representation, fang2023matching}. 
Various automated project recommendation tools have also been proposed, that match projects to developers with the closest technical background~\cite{zhang2017devrec}. 
Such matching is likely useful, as open-source projects depend heavily on contributions from volunteers, and contributors can be in short supply~\cite{coelho2017modern}.
At the same time, our results suggest that maximizing technical similarity between a project and the backgrounds of its contributors might be counterproductive in terms of enabling innovation.
More research is needed to explore this potential trade-off between \textit{exploration} (when divergent thinking is likely beneficial) and \textit{exploitation} (when narrow focus is required)~\cite{liu2021understanding} in a software development context, particularly open source.
For example, while tracking novelty computationally is becoming increasingly feasible (our work demonstrates this), we are still lacking in our understanding of when in the lifecycle of an open-source project innovations occur and who is responsible for them.
Future work could start exploring how novel design emerges at the onset of a project and how that compares to, \eg feature request discussions (by definition opportunities for innovation) occurring throughout a project's lifetime.

\mysec{Diversity}
Questions around the value of having access to diverse information have also come up in research focused on demographic team composition, \eg around gender~\cite{vasilescu2015gender} and race / ethnicity~\cite{shameer2023relationship}. 
This past work posits a similar underlying mechanism (surface-level demographic variables act as a proxy for deeper-level, less observable differences in backgrounds, skills, approaches to solve problems, etc) to the one we measure here more directly. 
Now that such measurements are becoming possible, it would be interesting to further test the relationships between team surface-level attributes, informational or network diversity, and outcomes.


\mysec{AI and Creativity}
Finally, we can't help but join colleagues~\cite{jackson2024creativity} in wondering how AI-based assistants will impact the innovativeness of the software being created. A large language model has certainly seen more diverse information in its training than any individual developer. But is the user experience designed to expose that diversity, \eg when generating code snippets automatically, or might AI-generated code end up looking more like a regression to the mean with current iteraction modalities?

\section{Conclusion}

Much like in Granovetter's jobs search study exposing the strength of weak ties, 
we conclude that open-source developers may also find useful ideas to help them create innovative projects through passive GitHub interactions, like starring repositories, more often than through active engagement like committing to a repository.



\mysec{Data Availability}
Our replication package\footnote{\url{https://github.com/icsesubmission/replication_package_icse26_novelty}} (Zenodo post acceptance) includes data and scripts to reproduce our tables and figures.



\balance
\bibliographystyle{ACM-Reference-Format}
\bibliography{references}

\newpage
\appendix

\section{Using transitivity to validate the `strength' of different networks}

\begin{figure}[ht]
\centering
\includegraphics[width=0.7\columnwidth, clip=true, trim=50 400 420 35]{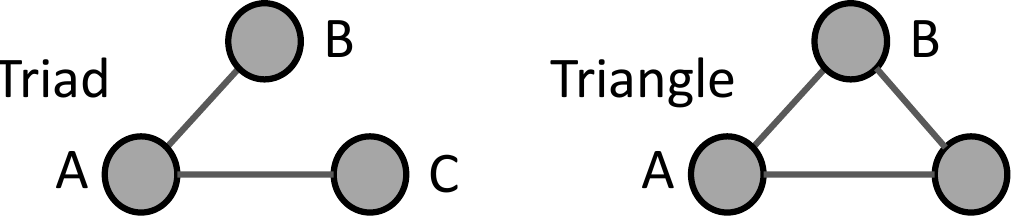}
\caption{In strongly tied social networks, triads are unlikely.}
\label{figure: triangles}
\end{figure}

In Section~\ref{sec:network-construction}, we reason that interactions via commits, issues, and stars reflect ties of decreasing strength, and here we provide an empirical validation of this assumption through the comparison of transitivity among the three networks.

Transitivity measures the density of local connections in the local network~\cite{newman2003structure}. 
Consider the illustration in Figure~\ref{figure: triangles}: if node $A$ is connected to $B$ and $C$, a triad is formed. 
In social network theory, if $A$'s connections to $B$ and $C$ are both strong ties, the triad is unstable and will likely evolve into a triangle where $A$, $B$, and $C$ are all connected. This evolution occurs because $B$ and $C$ have more opportunities to interact and increase their mutual familiarity due to their strong ties with $A$. 
Thus, networks characterized by strong ties typically exhibit a greater number of triangles, as opposed to triads, compared to those with weak ties~\cite{granovetter1973strength}. 
Formally, this can be computed using the measurement of transitivity, defined as $T = 3 * N_{\text{triangles}} / N_{\text{triads}}$.
A higher transitivity value (number of triangles relative to triads) indicates a network of stronger ties.

Similarly, we compare the transitivity values of our three networks, summarized in Table~\ref{table: tie strength}.\footnote{Since network transitivity is usually defined for undirected networks, we first convert our three networks (which are all directed) to their undirected form. That is, there is an undirected edge between nodes $A$ and $B$ if there is either an edge from $A$ to $B$ or from $B$ to $A$.
In addition, the reported number of nodes excludes isolates (\ie nodes without edges) from each network, and the number of edges is calculated in the undirected version of each network.}
Overall, we observe approximately an order of magnitude ($10 \times$) difference in transitivity values between each pair of networks.
Specifically, the commit network displays the highest levels of transitivity, followed by the issue network, while the star network exhibits the lowest level. Thus, these findings are consistent with our theoretical understanding of tie strength.

\begin{table}[ht]
\setlength{\tabcolsep}{1pt}
\caption{Validating the relatively decreasing strength of commit, issue, and star network ties.}
\label{table: tie strength}
  \centering \small
\begin{tabular}{P{1.4cm} D{)}{)}{9)3}@{} D{)}{)}{9)3}@{}  D{)}{)}{9)3}@{} } 
\toprule
 \multicolumn{1}{c}{\textbf{Interaction}} & \multicolumn{1}{c}{\textbf{\#Nodes}} & \multicolumn{1}{c}{\textbf{\#Edges}}  & \multicolumn{1}{c}{\textbf{Transitivity ($\times 10^{-2}$)}}\\
 \midrule
Commits           & 763,062  &  1,926,978  &  30.04 \\
Issues         &  278,945  & 727,255 & 3.42  \\
Stars       &  480,394   &  3,658,543 &  0.23 \\
\bottomrule

\end{tabular} 
\end{table}

\section{Replication with the atypicality measure by Fang et al}

We further test the robustness of our regression results by replicating the analysis presented in Table~\ref{table: regression analysis} using the original measurement of atypicality defined by \citet{fang2024novelty} as the outcome variable.
The results of this robustness analysis are reported in Table~\ref{table: regression analysis validation}.
It is important to note that the range of the atypicality score in the main analysis differs from that in the robustness analysis. As a result, the magnitudes of the estimated coefficients may vary; however, the overall qualitative patterns remain consistent.

\begin{table}[h] 
\centering \small
\renewcommand{\arraystretch}{0.8}
\setlength{\tabcolsep}{2pt}
  \caption{Regression analysis on factors associated with differences in project innovativeness, with innovativeness measured with the atypicality measure by \citet{fang2024novelty}}.
  \label{table: regression analysis validation} 
\begin{tabular}{P{3cm}cccc} 
\toprule
& Model I & Model II & Model III & Model IV\\ 
\midrule
  \textit{\textbf{Variables of interest}} & & & & \\
$\textit{Deg}_{ave}$ (\hr{hyp1}) & $-$0.055$^{***}$ & $-$0.009$^{*}$ &  & $-$0.017$^{***}$ \\ 
  & (0.0015) & (0.0044) &  & (0.0045) \\ 
  & & & & \\ 
$\textit{Deg}_{weakness}$ & 0.009$^{***}$ & 0.012$^{*}$ &  & $-$0.020$^{***}$ \\ 
  & (0.0022) & (0.0054) &  & (0.0059) \\ 
  & & & & \\ 
 $\textit{Div}_{ave}$ (\hr{hyp2}) &  &  & 0.042$^{***}$ & 0.048$^{***}$ \\ 
  &  &  & (0.0042) & (0.0044) \\ 
  & & & & \\ 
  $\textit{Div}_{weakness}$ (\hr{hyp3}) &  &  & 0.045$^{***}$ & 0.053$^{***}$ \\ 
  &  &  & (0.0056) & (0.0060) \\ 
  \textit{\textbf{Controls}} & & & & \\
 \textit{Org\_owned} & 0.125$^{***}$ & 0.071$^{***}$ & 0.074$^{***}$ & 0.076$^{***}$ \\ 
  & (0.0058) & (0.0118) & (0.0116) & (0.0118) \\ 
  & & & & \\ 
 $N_{\text{Owner\_Star}}$ (log)  & 0.016$^{***}$ & 0.011$^{***}$ & 0.010$^{***}$ & 0.013$^{***}$ \\ 
  & (0.0020) & (0.0028) & (0.0027) & (0.0028) \\ 
  & & & & \\ 
 $N_{\text{Core\_Dev}}$ (log) & 0.189$^{***}$ & 0.093$^{***}$ & 0.087$^{***}$ & 0.094$^{***}$ \\ 
  & (0.0053) & (0.0114) & (0.0112) & (0.0114) \\ 
  & & & & \\ 
 $N_{\text{Packages}}$ (log) & 0.236$^{***}$ & 0.150$^{***}$ & 0.153$^{***}$ & 0.154$^{***}$ \\ 
  & (0.0031) & (0.0080) & (0.0079) & (0.0080) \\ 
 & & & & \\ 
 \textit{\textbf{Fixed effect}} & & & & \\
$\text{Year}_{\text{creation}}$ & \checkmark & \checkmark & \checkmark & \checkmark \\ 
\midrule
Observations & 589,999 & 37,451 & 37,451 & 37,451 \\ 
Adjusted R$^{2}$ & 0.044 & 0.034 & 0.038 & 0.038 \\ 
\bottomrule
&   \multicolumn{4}{r}{$^{*}$p$<$0.05; $^{**}$p$<$0.01; $^{***}$p$<$0.001} \\ 
\end{tabular} 
\end{table}


\section{Replication with Alternative Definitions of Core Developers and Interaction Periods}

In Section~\ref{sec:network-construction}, we constructed knowledge networks between projects based on interactions among their \textit{core developers} within a \textit{12-month} period prior to their first contributions.
To assess the robustness of our findings, we conducted replication analyses using alternative definitions of core developers and varying lengths of interaction periods, as reported in Table~\ref{table: regression analysis}.

Following established practice~\cite{mockus2002two}, we alternatively identify core developers as the top contributors to a project whose cumulative commits account for at least 80\% of all project commits. The corresponding regression results using this alternative definition are presented in Table~\ref{table: regression analysis validation, alternative core}.
Similarly, we extend the interaction window from 12 to 24 months prior to a core developer’s first contribution to the focal project when constructing the networks. The regression results based on this extended interaction period are presented in Table~\ref{table: regression analysis validation, 24-month period}.

We also run the same model (as model IV in Table~\ref{table: regression analysis}) with several other combinations of core developer identification approach and length of interaction periods, and the estimated coefficient of Variables of Interest are presented in Table~\ref{table:regression_validation_combined}

\begin{table}[H]  
\centering \small
\renewcommand{\arraystretch}{0.6}
\setlength{\tabcolsep}{2pt}
  \caption{Regression analysis with alternative core developer identification approach} 
  \label{table: regression analysis validation, alternative core} 
\vspace{-0.2cm}
\begin{tabular}{P{3cm}cccc} 
\toprule
& Model I & Model II & Model III & Model IV\\ 
\midrule
  \textit{\textbf{Variables of interest}} & & & & \\
$\textit{Deg}_{ave}$ (\hr{hyp1}) & 0.001$^{***}$ & $-$0.001 &  & $-$0.002$^{***}$ \\ 
  & (0.0002) & (0.0006) &  & (0.0006) \\ 
  & & & & \\ 
$\textit{Deg}_{weakness}$ & $-$0.001$^{**}$ & $-$0.002$^{*}$ &  & $-$0.006$^{***}$ \\ 
  & (0.0002) & (0.0007) &  & (0.0008) \\ 
  & & & & \\ 
 $\textit{Div}_{ave}$ (\hr{hyp2}) &  &  & 0.007$^{***}$ & 0.008$^{***}$ \\ 
  &  &  & (0.0006) & (0.0006) \\ 
  & & & & \\ 
  $\textit{Div}_{weakness}$ (\hr{hyp3}) &  &  & 0.004$^{***}$ & 0.006$^{***}$ \\ 
  &  &  & (0.0007) & (0.0008) \\ 
  \textit{\textbf{Controls}} & & & & \\
 \textit{Org\_owned} & 0.028$^{***}$ & 0.018$^{***}$ & 0.020$^{***}$ & 0.020$^{***}$ \\ 
  & (0.0006) & (0.0016) & (0.0015) & (0.0016) \\ 
  & & & & \\ 
 $N_{\text{Owner\_Star}}$ (log)  & 0.002$^{***}$ & 0.001 & 0.0004 & 0.0007 \\ 
  & (0.0002) & (0.0004) & (0.0004) & (0.0004) \\ 
  & & & & \\ 
 $N_{\text{Core\_Dev}}$ (log) & 0.010$^{***}$ & 0.017$^{***}$ & 0.015$^{***}$ & 0.016$^{***}$ \\ 
  & (0.0005) & (0.0015) & (0.0015) & (0.0015) \\ 
  & & & & \\ 
 $N_{\text{Packages}}$ (log) & 0.043$^{***}$ & 0.025$^{***}$ & 0.025$^{***}$ & 0.026$^{*}$ \\ 
  & (0.0003) & (0.0010) & (0.0010) & (0.0010) \\ 
 & & & & \\ 
 \textit{\textbf{Fixed effect}} & & & & \\
$\text{Year}_{\text{creation}}$ & \checkmark & \checkmark & \checkmark & \checkmark \\ 
\midrule
Observations & 679,930 & 40,589 & 40,589 & 40,589 \\ 
Adjusted R$^{2}$ & 0.053 & 0.064 & 0.068 & 0.070 \\ 
\bottomrule
&   \multicolumn{4}{r}{$^{*}$p$<$0.05; $^{**}$p$<$0.01; $^{***}$p$<$0.001} \\ 
\end{tabular} 
\vspace{-0.5cm}
\end{table}

\begin{table}[H] 
\centering \small
\renewcommand{\arraystretch}{0.6}
\setlength{\tabcolsep}{2pt}
  \caption{Regression analysis with 24-month period length to construct networks} 
  \label{table: regression analysis validation, 24-month period} 
\vspace{-0.2cm}
\begin{tabular}{P{3cm}cccc} 
\toprule
& Model I & Model II & Model III & Model IV\\ 
\midrule
  \textit{\textbf{Variables of interest}} & & & & \\
$\textit{Deg}_{ave}$ (\hr{hyp1}) & 0.002$^{***}$ & 0.0003 &  & $-$0.001$^{**}$ \\ 
  & (0.0002) & (0.0005) &  & (0.0005) \\ 
  & & & & \\ 
$\textit{Deg}_{weakness}$ & $-$0.0002 & $-$0.002$^{**}$ &  & $-$0.007$^{***}$ \\ 
  & (0.0002) & (0.0006) &  & (0.0007) \\ 
  & & & & \\ 
 $\textit{Div}_{ave}$ (\hr{hyp2}) &  &  & 0.007$^{***}$ & 0.008$^{***}$ \\ 
  &  &  & (0.0004) & (0.0005) \\ 
  & & & & \\ 
  $\textit{Div}_{weakness}$ (\hr{hyp3}) &  &  & 0.004$^{***}$ & 0.007$^{***}$ \\ 
  &  &  & (0.0007) & (0.0007) \\ 
  \textit{\textbf{Controls}} & & & & \\
 \textit{Org\_owned} & 0.027$^{***}$ & 0.020$^{***}$ & 0.022$^{***}$ & 0.021$^{***}$ \\ 
  & (0.0006) & (0.0014) & (0.0014) & (0.0014) \\ 
  & & & & \\ 
 $N_{\text{Owner\_Star}}$ (log)  & 0.002$^{***}$ & 0.001$^{*}$ & 0.0010$^{**}$ & 0.0012$^{***}$ \\ 
  & (0.0002) & (0.0003) & (0.0003) & (0.0003) \\ 
  & & & & \\ 
 $N_{\text{Core\_Dev}}$ (log) & 0.013$^{***}$ & 0.015$^{***}$ & 0.015$^{***}$ & 0.015$^{***}$ \\ 
  & (0.0006) & (0.0014) & (0.0013) & (0.0013) \\ 
  & & & & \\ 
 $N_{\text{Packages}}$ (log) & 0.042$^{***}$ & 0.024$^{***}$ & 0.025$^{***}$ & 0.025$^{***}$ \\ 
  & (0.0003) & (0.0009) & (0.0009) & (0.0009) \\ 
 & & & & \\ 
 \textit{\textbf{Fixed effect}} & & & & \\
$\text{Year}_{\text{creation}}$ & \checkmark & \checkmark & \checkmark & \checkmark \\ 
\midrule
Observations & 587,710 & 51,571 & 51,571 & 51,571 \\ 
Adjusted R$^{2}$ & 0.055 & 0.055 & 0.059 & 0.061 \\ 
\bottomrule
&   \multicolumn{4}{r}{$^{*}$p$<$0.05; $^{**}$p$<$0.01; $^{***}$p$<$0.001} \\ 
\end{tabular} 
\vspace{-0.5cm}
\end{table}

\begin{table}[H]
\centering \small
\renewcommand{\arraystretch}{0.7}
\setlength{\tabcolsep}{3pt}
\caption{Regression analysis with different core developer identification and length of interaction periods}
\label{table:regression_validation_combined}
\vspace{-0.2cm}

\begin{tabular}{>{\centering\arraybackslash}p{1.5cm} >{\centering\arraybackslash}p{1.5cm} cccc} 
\toprule
 & & \multicolumn{4}{c}{\textbf{Variables}} \\  
\cmidrule(lr){3-6}
\parbox[t]{1.5cm}{\raggedright\textbf{Core \\developer\\ identifica\\-tion}} &
\parbox[t]{1.5cm}{\raggedright\textbf{Interaction \\period length\\ (month)}} &
$\textit{Deg}_{ave}$ & $\textit{Deg}_{weakness}$ & $\textit{Div}_{ave}$ & $\textit{Div}_{weakness}$ \\
\midrule

Original & 6 &  $-$0.001 & $-$0.005$^{***}$ & 0.006$^{***}$ & 0.006$^{***}$ \\
 &  & (0.0009) & (0.0011) & (0.0008) & (0.0008) \\
 &  & & & & \\

Alternative & 6 &  $-$0.003$^{**}$ & $-$0.006$^{***}$ & 0.007$^{***}$ & 0.007$^{***}$ \\
 &  & (0.0008) & (0.0010) & (0.0008) & (0.0010) \\
 &  & & & & \\

Alternative & 24 &  $-$0.003$^{**}$ & $-$0.007$^{***}$ & 0.010$^{***}$ & 0.006$^{***}$ \\
 &  & (0.0005) & (0.0007) & (0.0005) & (0.0007) \\
 
\bottomrule
& \multicolumn{5}{r}{$^{*}$p$<$0.05; $^{**}$p$<$0.01; $^{***}$p$<$0.001} \\
\end{tabular}
\vspace{-0.4cm}
\end{table}

\section{Relationship between awesome and atypical projects.}
\label{appendix: Awesome atypical validation}

\begin{table}[h] 
\centering \small
\renewcommand{\arraystretch}{0.8}
\setlength{\tabcolsep}{2pt}
\caption{Regression analysis on the relationship between awesome projects and project innovativeness. (Outcome variable: Project Atypicality)} 
\label{table:regression_atypicality_innovation} 
\begin{tabular}{P{3cm}c} 
\toprule
 & Estimated coefficient \\ 
\midrule
\textit{\textbf{Variables of interest}} & \\
\textit{Is awesome}  & 0.028$^{***}$ \\
 & (0.0052) \\
\textit{\textbf{Controls}} & \\
\textit{Org\_owned} & 0.017$^{***}$ \\
 & (0.0016) \\
$N_{\text{Owner\_Star}}$ (log) & 0.001 \\
 & (0.0004) \\
$N_{\text{Core\_Dev}}$ (log) & 0.017$^{***}$ \\
 & (0.0015) \\
$N_{\text{Packages}}$ (log) & 0.023$^{***}$ \\
 & (0.0011) \\
\textit{\textbf{Fixed effect}} & \\
$\text{Year}_{\text{creation}}$ & \checkmark \\
\midrule
Observations & 37,451 \\
Adjusted R$^{2}$ & 0.059 \\ 

\bottomrule
& \multicolumn{1}{r}{$^{*}$p$<$0.05; $^{**}$p$<$0.01; $^{***}$p$<$0.001} \\ 
\end{tabular} 
\end{table}

In Section~\ref{validation awesome-atypicality}, we present that the ``awesome'' projects on average are more atypical per our measurement than other projects in our sample, and we further validate this result with a regression analysis in Table~\ref{table:regression_atypicality_innovation}, where we show that this relationship is not driven by confounding variables such as the project size and team size.

\section{Real-world examples of projects with varying level of tie diversity}

\begin{figure*}[h]
\centering
\includegraphics[width=0.9\linewidth, clip=true]{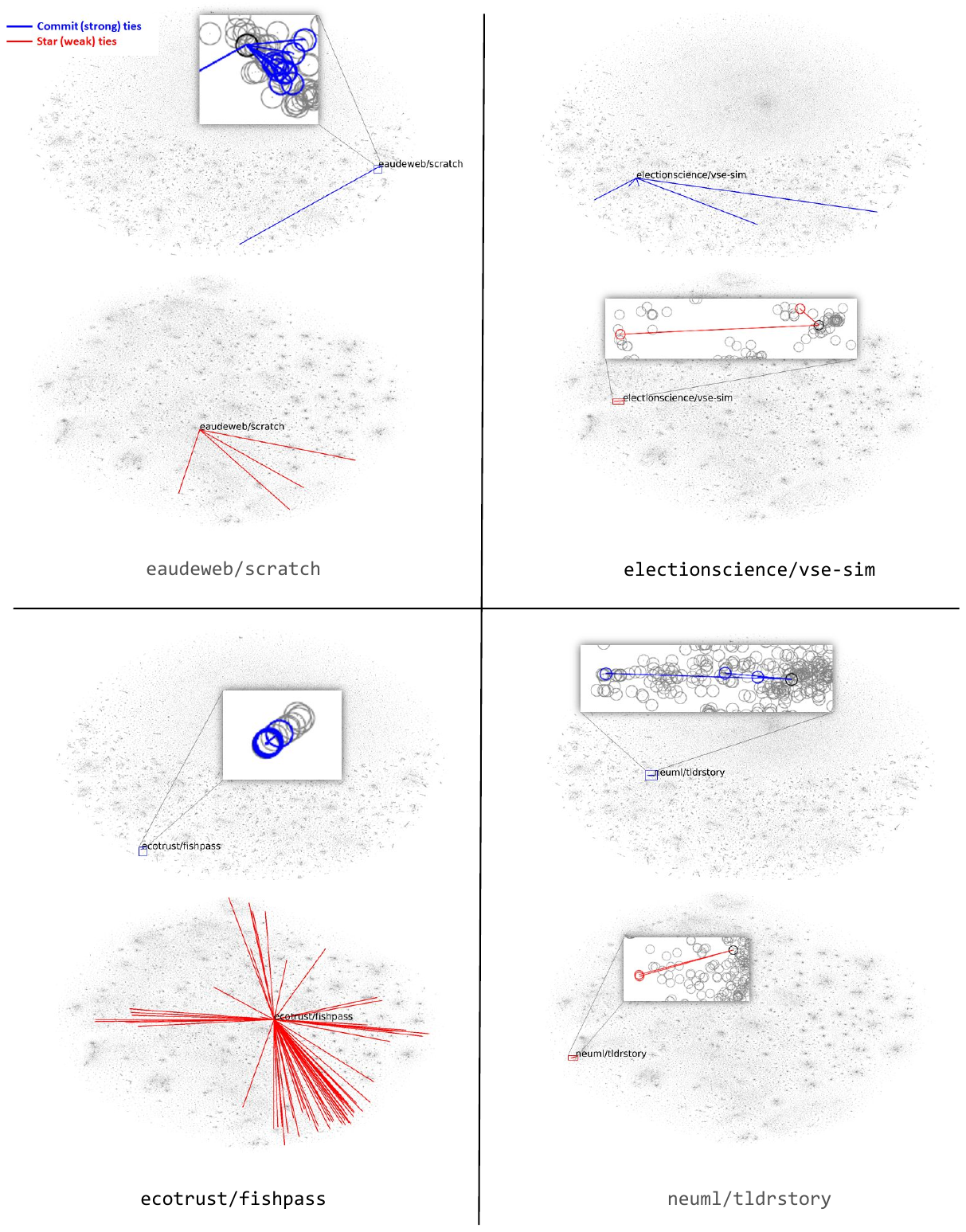}
\caption{Real-world examples for projects with varying level of diversity, corresponding to the four quadrants in Figure~\ref{figure:pca-ties}.}
\label{figure:real-world}
\end{figure*}

In Section~\ref{subsec: PCA}, we present graphs depicting the networks of projects characterized by varying levels of average diversity and relative weak tie diversity. To concretize those abstract diagrams, we also provide real-world examples in Figure~\ref{figure:real-world}.

\end{document}